\documentclass[sigconf,natbib=false,nonacm=true]{acmart}

\AtBeginDocument{%
  }

\RequirePackage[
  datamodel=acmdatamodel,
  style=acmnumeric,
  url=true,
  doi=false,
  isbn=false
]{biblatex}

\usepackage{subcaption}
\usepackage{multirow}
 
\begin{document}

\title{Behavior Change as a Signal\\for Identifying Social Media Manipulation}

\author{Isuru Ariyarathne}
\orcid{0000-0003-1811-7680}
\affiliation{
  \institution{Department of Data Science}
  \institution{William \& Mary}
  \city{Williamsburg}
  \state{Virginia}
  \country{USA}
}
\email{iahewababarand@wm.edu}

\author{Gangani Ariyarathne}
\orcid{0000-0003-4205-6574}
\affiliation{
  \institution{Department of Data Science}
  \institution{William \& Mary}
  \city{Williamsburg}
  \state{Virginia}
  \country{USA}
}
\email{gchewababarand@wm.edu}

\author{Alessandro Flammini}
\orcid{0000-0003-1670-9156}
\affiliation{%
  \institution{Observatory on Social Media\\Indiana University}
  \city{Bloomington}
  \state{Indiana}
  \country{USA}
}
\email{aflammin@iu.edu}

\author{Filippo Menczer}
\orcid{0000-0003-4384-2876}
\affiliation{%
  \institution{Observatory on Social Media\\Indiana University}
  \city{Bloomington}
  \state{Indiana}
  \country{USA}
}
\email{fil@iu.edu}

\author{Alexander C. Nwala}
\orcid{0000-0003-3408-791X}
\affiliation{
  \institution{Department of Data Science}
  \institution{William \& Mary}
  \city{Williamsburg}
  \state{Virginia}
  \country{USA}
}
\email{acnwala@wm.edu}

\begin{abstract}
Social media accounts engaging in online manipulation can change their behaviors for re-purposing or to evade detection. 
Existing detection systems are built on features that do not exploit such behavioral patterns. 
Here we investigate the degree to which change in behavior can serve as a signal for identifying automated or coordinated accounts. 
First, we use Behavioral Languages for Online Characterization (BLOC) to represent the behavior of a social media account as a sequence of symbols that represent the account's actions and content. 
Second, we segment an account's BLOC strings and measure the changes between consecutive segments. 
Third, we represent an account as a feature vector that captures the distribution of behavioral change values. 
Finally, the resulting features are used to train and test supervised classifiers. 
We apply the proposed method to two detection tasks aimed at automated behavior (social bots) and coordinated inauthentic behavior (information operations).  
Our results reveal that the distributions of behavioral changes tend to be consistent across authentic accounts, while social bots exhibit either very low or very high behavioral change. 
Coordinated inauthentic accounts exhibit highly similar distributions of behavioral change within the same campaign, but diverse across campaigns. 
These patterns allow our classifiers to achieve good accuracy in both tasks, demonstrating the effectiveness of behavioral change as a signal for identifying online manipulation.
\end{abstract}


\maketitle

\section{Introduction}

Abusive accounts on social media, like inauthentic automated (bot) or coordinated accounts, have received significant research attention due to the threats they pose to the integrity of online platforms for spreading misinformation~\cite{fake_news_spreader}, manipulating public opinion~\cite{2016_presidential_election, bessi2016social, deverna2025modeling}, and perpetrating fraud~\cite{cresci2019cashtag}. Despite many advances, identifying these deceptive accounts remains a challenging task due to their ability to evolve and adapt~\cite{decade_of_social_bots, cresci2023demystifying, yang2023social}.

Supervised detection systems are usually built from machine-learning models that are trained on features capturing some aspects of account behaviors. Features can be derived from the content of posts~\cite{dickerson2014using}, user metadata~\cite{ali2023real}, network structure~\cite{pacheco2021uncovering, rodriguez2020one,smith2025unsupervised}, and so on. While useful, existing approaches trained on such features are typically designed to detect specific types of suspicious behaviors. Since malicious actors adjust their tactics to appear normal, switch purposes, and avoid detection, such approaches do not generalize well~\cite{yang2020scalable,sayyadiharikandeh2020detection,cresci2023demystifying}. For example, Cresci et al. \cite{Cresci_Petrocchi_Spognardi_Tognazzi_2021} demonstrated how coordinated accounts evolve their activity patterns over time to resemble legitimate users. Similarly, Elmas et al. \cite{elmas2022misleadingrepurposingtwitter} documented large scale account repurposing were accounts abruptly changed identity and intent. These findings suggest that behavioral change itself is a recurring tactic in online manipulation.
However, behavioral change is not exclusive to malicious actors, as genuine users also naturally evolve over time. Moreover, existing methods are less adaptable in adversarial settings not because their features are incorrect, but because inauthentic accounts strategically shift behavior in ways that cause those signals to weaken or disappear. Therefore, it is important to examine whether behavioral change can be systematically used to distinguish manipulation from normal user evolution.

\begin{figure*}
\centering
\includegraphics[width=0.9\textwidth]{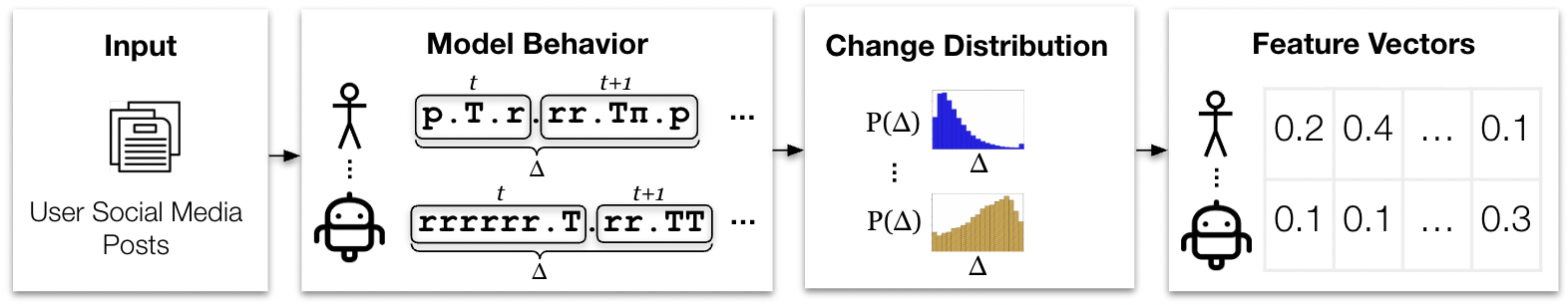}
\caption{Overview of our Behavior Change methodology to identify social media manipulation. The feature vectors are used to train a model for a specific task, such as detecting automated or coordinated accounts.}
\label{fig:overview}
\end{figure*}

We use the term \textit{behavioral change} or \textit{behavior change} to describe how online user action and content patterns differ across consecutive time spans (Section~\ref{sec:measure_change} provides a more formal definition). Our approach (Fig.~\ref{fig:overview}) begins by representing social media accounts with Behavioral Languages for Online Characterization (BLOC)~\cite{nwala_flammini_menczer_bloc}. BLOC encodes the activities of a social media account as strings of symbols drawn from alphabets that represent the account's actions or the content of its posts. (BLOC could be replaced by another representation that captures actions and content behaviors.) 
We segment an account's BLOC words and measure the distance (change) between consecutive segments. 
We investigate multiple segmentation methods, segment selection methods, and distance measures as discussed in detail in Section~\ref{sec:measuring_change}. 
We train detectors of bot and coordinated accounts using features that capture the resulting distributions of behavioral change values. 

Our analysis shows that organic accounts tend to have normal behavioral change patterns, exhibiting neither very repetitive behaviors nor very drastic changes. Conversely, some bot accounts exhibit unusually small change, suggesting automated behavior, while others display drastic or erratic changes, possibly when they are repurposed or to avoid detection. Coordinated inauthentic accounts belonging to the same campaign exhibit highly similar behavioral changes, signaling a lack of independence. 
We demonstrate the usefulness of these signals by developing supervised machine learning models that leverage behavioral change features to identify the two types of online manipulation. 
These results suggest that behavioral change could be leveraged to detect other forms of online abuse. We have published our code to encourage this: \textcolor{blue}{https://github.com/wm-newslab/behavior-change}.

\section{Related Studies}

Social media abuse takes different forms such as cyberbullying, fraud, disinformation, impersonation using deep fakes, and so on~\cite{thomas2013trafficking, nauta2017detecting, farazmanesh2022compromised, yen2021detecting, cresci2019cashtag, diaz2025survey}. 
Since abusers often rely on automated or coordinated accounts to increase the effectiveness of their attacks, these kinds of social media abuse have received significant research attention. Predominant strategies for detecting both inauthentic automated and coordinated accounts rely on supervised and unsupervised machine-learning models to distinguish them from genuine ones~\cite{decade_of_social_bots, pacheco2021uncovering}. 

Automation detection approaches utilize classes of features that include metadata such as tweet frequency~\cite{chu2012detecting}, profile descriptions~\cite{ali2023real}, follower-friend ratios~\cite{gilani2017depth}, and network structure~\cite{cao2014uncovering}, as well as content-based attributes like lexical patterns~\cite{inuwa2018lexical} and hashtag or URL usage~\cite{varol2017online}. While these methods have proven successful in detecting some classes of bots, they struggle due to the evolving sophistication of malicious bots. 
Bots can now use LLMs to generate natural language, mimic human behavior, build elaborate social networks, and interact with genuine users online~\cite{anatomyBotnet, wack2025generative}. This makes simple heuristics like follower counts and posting rates unreliable, leading to detection errors \cite{ferrara2016rise, decade_of_social_bots, cresci2023demystifying}.

In coordination detection, researchers aim to uncover suspiciously similar behaviors among groups of accounts, such as identical retweet patterns or synchronized posts. Network based models and clustering algorithms are often used for identifying coordinated activities on social media~\cite{cao2014uncovering, pacheco2021uncovering}. Recent studies show that coordinated inauthentic accounts adjust their posting intervals~\cite{weber2022temporalc} and modify their language to imitate genuine communities~\cite{socioLinguisticCia}, making coordinated behavior harder to detect. This ongoing evolution reflects an arms race with detection systems and highlights the need for more advanced approaches.

Behavioral modeling is a promising approach in the latest generation of detection systems. Models such as Act-M~\cite{costa2017modeling}, MulBot~\cite{mannocci2022mulbot}, BotShape~\cite{wu2023botshape}, and Temporal Network Logic~\cite{pedersen2023detecting} identify anomalies by examining posting intervals, changes across behavioral dimensions, and transitions in user activities over time. Moreover, methods such as Digital DNA~\cite{cresci2016dna} and the BLOC framework~\cite{nwala_flammini_menczer_bloc} encode user behaviors as sequences of symbols to reveal repetitive or coordinated patterns. 

\section{Measuring Change in the Behaviors of Social Media Accounts}
\label{sec:measuring_change}

Building upon prior behavioral models, we focus on behavioral change as a signal for identifying suspicious behaviors rather than relying on features that capture predefined abusive behaviors. 
When examining whether an account behaves like known abusive bots or coordinated actors, rather than focusing on features about the behavior itself, we inspect how such behavior changes over time.  For example, an account that suddenly becomes highly active after a long dormant period, or an account that shifts from sharing personal content to spreading links to external websites, may flag suspicious behavioral anomalies even if they do not match any previously-identified abuse pattern.  

Before explaining our approach for measuring change in the behaviors of social media accounts, let us first explain the BLOC model we used for representing behaviors. As discussed in the introduction, one could use a different model to represent behaviors. 

\subsection{BLOC Framework}

\begin{figure}
    \centering
    \includegraphics[width=0.47\textwidth]{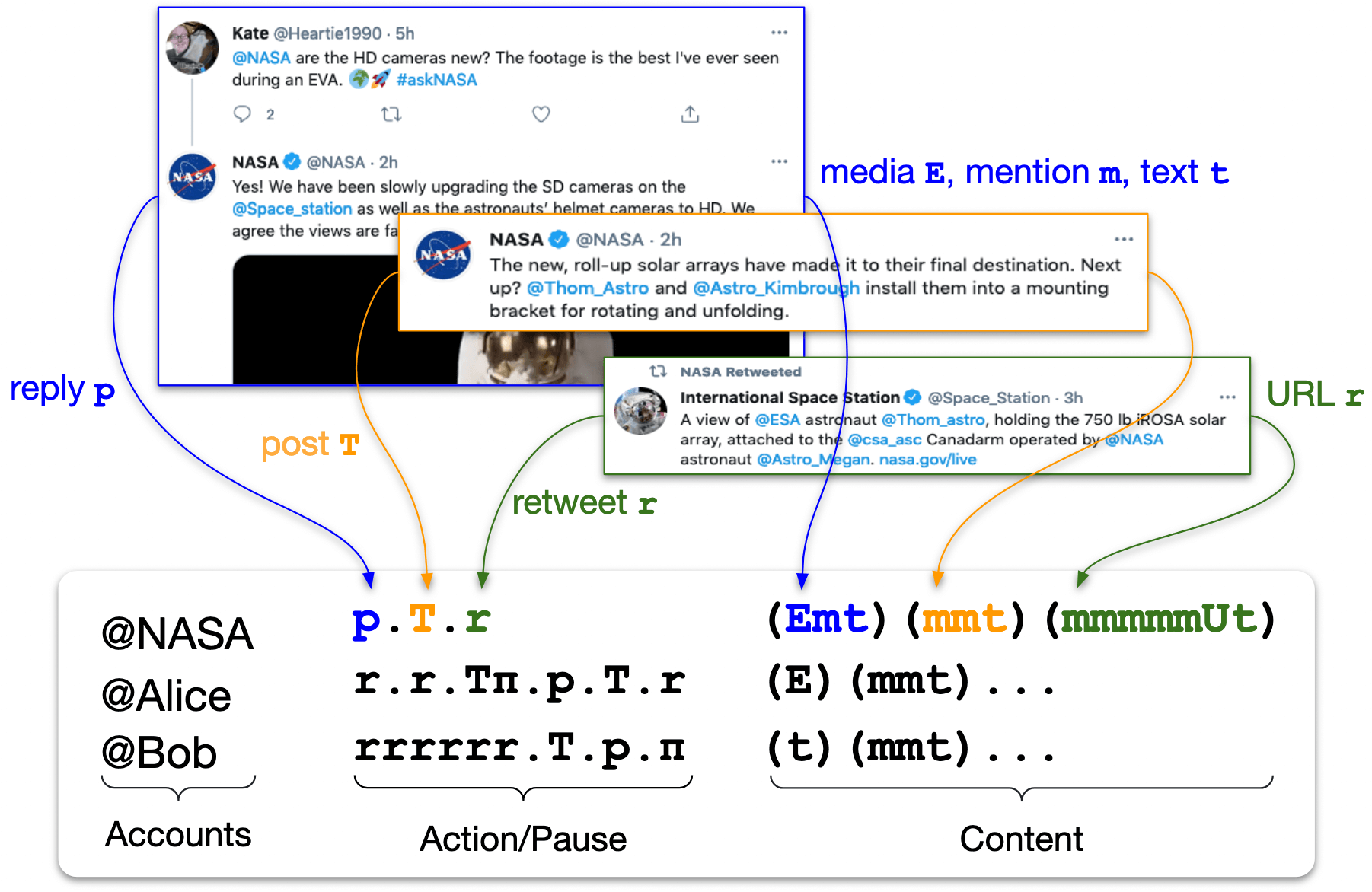} 
    \caption{BLOC representation of the behaviors of three users, \texttt{@NASA}, \texttt{@Alice}, and \texttt{@Bob} using action/pause and content alphabets. For the action/pause alphabet, the sequence of three tweets (a reply, an original tweet, and a retweet) by \texttt{@NASA} can be represented by three letters $p.T.r$ separated by dots (long pauses). Using the content alphabet, it can be represented by three sets of strings $(Emt)(mmt)(mmmmmUt)$ enclosed in parentheses.}
    \label{fig:BLOC_example}
\end{figure}

The BLOC (Behavioral Languages for Online Characterization) framework~\cite{nwala_flammini_menczer_bloc} provides formal languages that can be used to represent the behaviors of social media accounts, irrespective of specific social media platforms, account types (human or bot), or intent (malicious or benign). 
BLOC strings consist of symbols drawn from distinct alphabets representing an account's \textit{actions} and \textit{content}.  
The BLOC \textit{action/pause} alphabet encodes user actions such as posting/tweeting ($T$), resharing/retweeting ($r$), and replying ($p$). 
Fig.~\ref{fig:BLOC_example} illustrates a possible representation of a sequence of tweets by three different Twitter\footnote{While the platform is now called X, in this paper we use Twitter because our training and evaluation data predates the name change.} accounts, \texttt{@NASA}, \texttt{@Alice}, and \texttt{@Bob}. 
The \texttt{@NASA} account replied to a tweet, posted a tweet, and then retweeted a tweet, resulting in the BLOC action sequence: $p.T.r$. 
Dots represent long pauses (e.g., more than one minute) between consecutive actions. 
In Fig.~\ref{fig:BLOC_example}, the same dot symbol ($.$) is used to represent different long pauses. However, BLOC enables the use of different symbols to distinguish different long pauses (hour, day, week, month, year). 
The \textit{content} alphabet encodes different kinds of post content, such as text ($t$), links ($U$), hashtags ($H$), media ($E$), and mentions ($M$), as also illustrated in Fig.~\ref{fig:BLOC_example}. 

\subsection{Measuring Change in Behavior}
\label{sec:measure_change}

\begin{figure*}
  \centering
  \begin{subfigure}{0.319\textwidth}
    \centering
    \includegraphics[width=\linewidth]{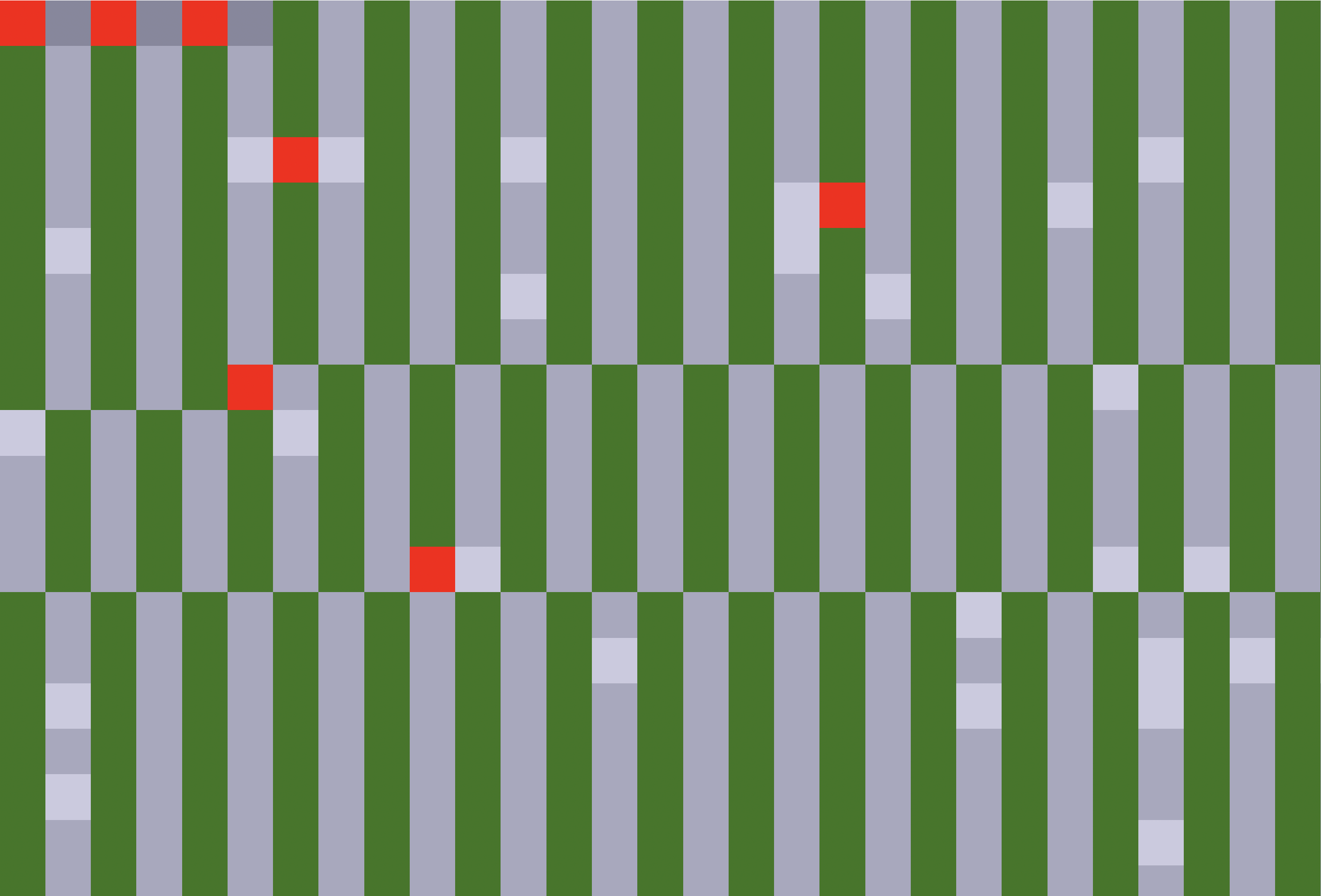}
    \caption{\texttt{@FoxNews}}
  \end{subfigure}
  \begin{subfigure}{0.32\textwidth}
    \centering
    \includegraphics[width=\linewidth]{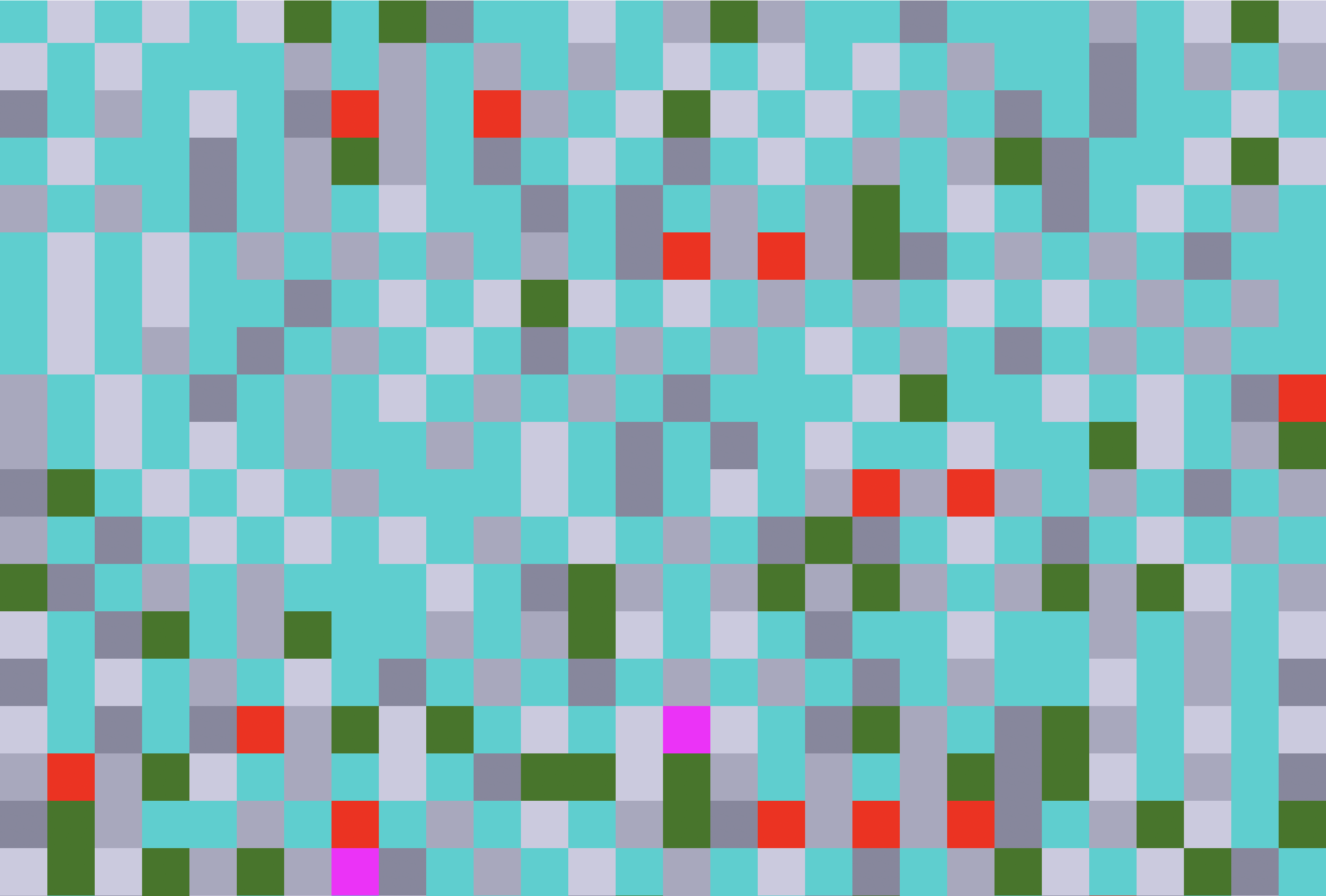}
    \caption{\texttt{@elonmusk}}
  \end{subfigure}
  \begin{subfigure}{0.33\textwidth}
    \centering
    \includegraphics[width=\linewidth]{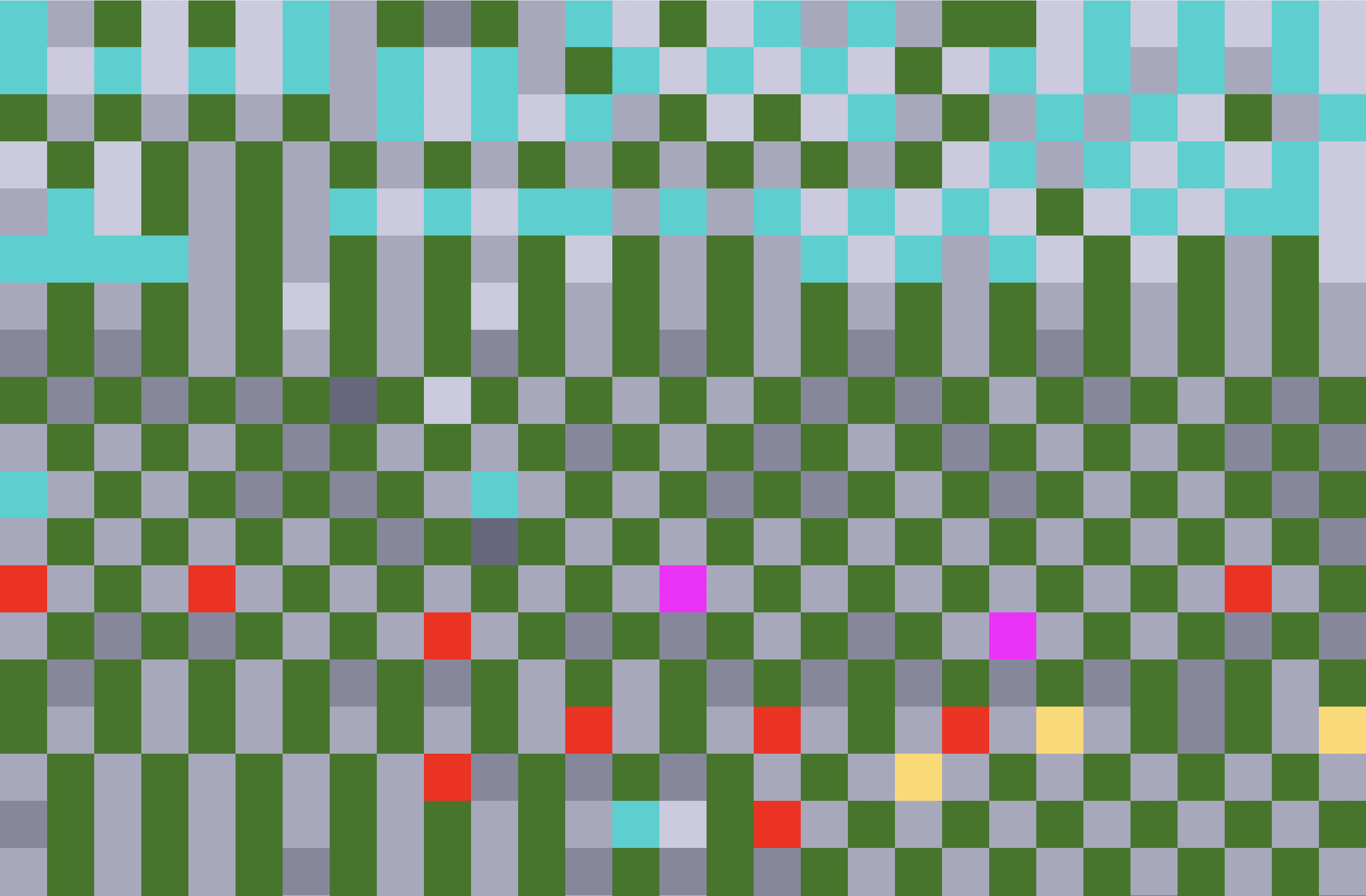}
    \caption{\texttt{@TEN\_GOP}}
  \end{subfigure}
  \caption{Color-coding of BLOC action symbols of Twitter accounts to illustrate how their behaviors change (color switches). Each square represents an action. Color legend (non-exhaustive): Green - \textit{post}, Red - \textit{retweet}, Cyan - \textit{reply}, Gray - pauses (Darker gray - longer pauses).
    (a)~\texttt{@FoxNews} exhibits repetitive patterns, reflecting automated activity.
    (b)~\texttt{@elonmusk} exhibits a mix of diverse actions (replies and posts) without repetitive patterns.
    (c)~\texttt{@TEN\_GOP}, a Russian troll account, exhibits sudden shifts between organic-looking and repetitive patterns.
    }
  \label{fig:BLOC-change}
\end{figure*}

\begin{figure}
    \centering
    \includegraphics[width=0.47\textwidth]{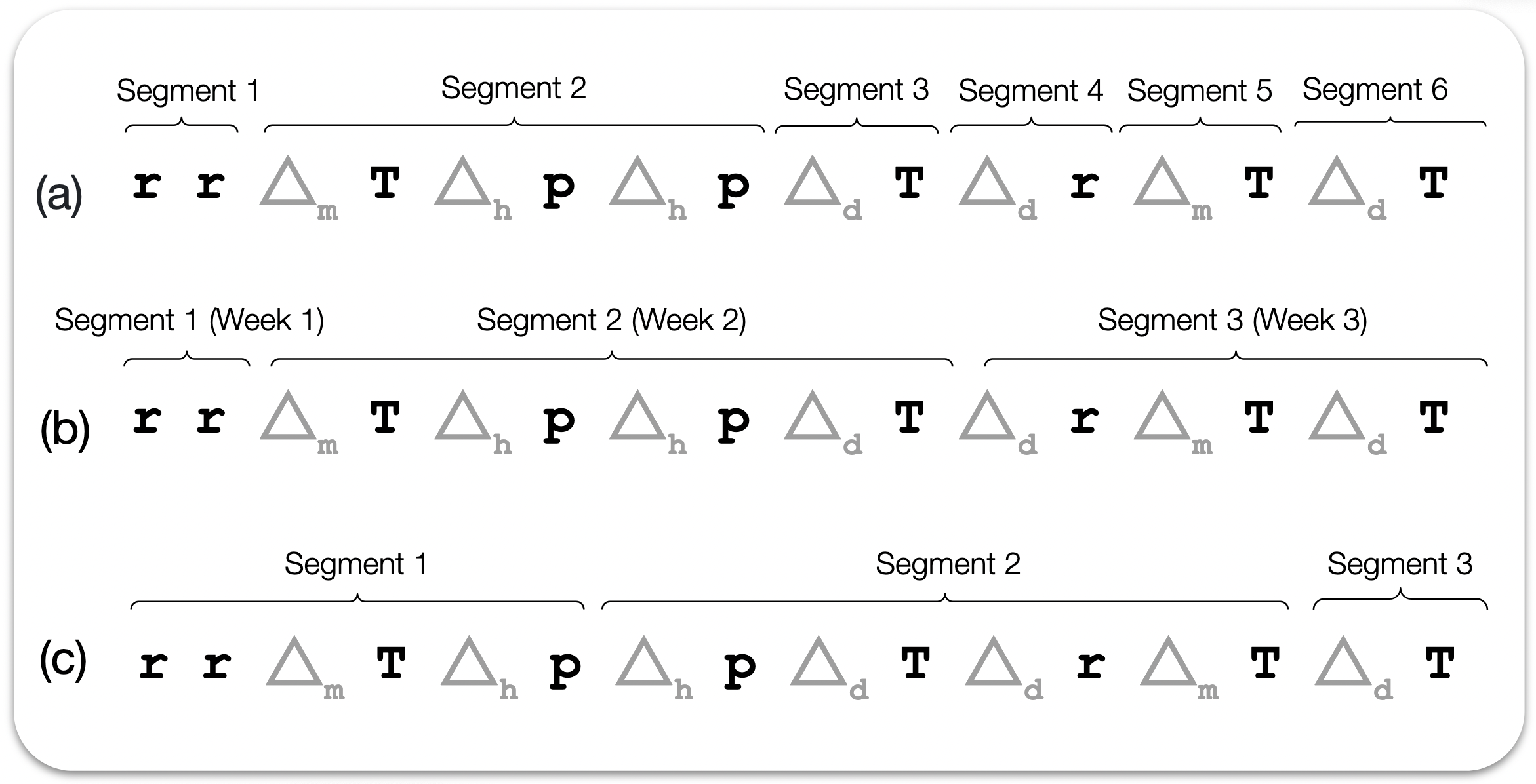} 
    \caption{Three ways of segmenting a user's BLOC string. 
    This example includes pauses less  than an hour ($\Delta_h$), between an hour and a day ($\Delta_d$), and between a day and a month ($\Delta_m$). 
    (a)~\textit{pauses} (longer than one hour). 
    (b)~\textit{weeks}. 
    (c)~\textit{sets-of-k} ($k=4$).
    }
    \label{fig:Segmentation}
\end{figure}

The BLOC strings in Fig~\ref{fig:BLOC-change} illustrate the behavioral patterns exhibited by different accounts. By analyzing these patterns and how they change over time, we can differentiate between authentic, and inauthentic behaviors. 
To capture such behavioral change patterns, we can segment BLOC strings into smaller partitions and compare the segments.  
We explored three segmentation methods as shown in Fig.~\ref{fig:Segmentation}. 
In segmentation by \textit{pauses}, a new segment starts after any period of inactivity longer than a predefined threshold (e.g., one hour). Each segment then represents a session of user activity. 
In segmentation by \textit{week}, each segment represents activities that occurred in the same week. 
These two approaches could result in BLOC segments of different lengths. 
Alternatively, segmentation by \textit{sets-of-k} partitions the string into sets of equal length $k$. 
While segmentation is based on action strings, for each action segment, we also create a content segment that includes the corresponding content symbols. 

After a user's behavior is partitioned into sequential segments, we must select different segments to be compared for measuring how behavior evolves over time. 
We explored two selection methods.
Consider a sequence of four segments $s_1$, $s_2$, $s_3$, and $s_4$ for illustration. In \textit{adjacent} selection, we compare consecutive segments, e.g., ($s_1$ vs. $s_2$), ($s_2$ vs. $s_3)$, and ($s_3$ vs. $s_4$). This method is sensitive to short-term fluctuations in behaviors. In \textit{cumulative} selection, we compare each segment with the concatenation of all prior segments, e.g., ($s_1$ vs. $s_2$), ($s_1 s_2$ vs. $s_3$), and ($s_1 s_2 s_3$ vs. $s_4$). This method captures how present behavior diverges from historical ones. 

We measure how an account's behavior evolves over time by computing the differences between selected pairs of segments. We propose two methods to measure these differences. For \textit{cosine} distance, we convert two selected segments into term frequency vectors $v_1, v_2$ \cite{nwala_flammini_menczer_bloc} and then compute their cosine distance ($1 - \cos(v_1, v_2)$). For \textit{compression} distance, we employ the \emph{Normalized Compression Distance (NCD)} \cite{cilibrasi2005clustering}. 
Let $C(\cdot)$ denote the compressed length (in bytes) of a string under a given 
compressor (e.g., zlib). The NCD between $s_1$ and $s_2$ is defined as
\[
\text{NCD}(s_1, s_2) = 
\frac{C(s_1 s_2) - \min \{ C(s_1), \, C(s_2) \}}
     {\max \{ C(s_1), \, C(s_2) \}} \, ,
\]
where $C(s_1 s_2)$ denotes the compressed length of the concatenation of 
$s_1$ and $s_2$. Intuitively, NCD measures the degree to which the two segments share compressible patterns. 
Both distance measures are defined in the unit interval.  

We summarize the key design choices through a triple that we call \textit{change setting}: (i)~the segmentation method (\textit{pauses}, \textit{weeks}, or \textit{sets of k}), (ii)~the segment selection method (\textit{adjacent}, \textit{cumulative}), and (iii)~the distance measure (\textit{cosine} or \textit{compression} distance). 

The set of distance values computed across all selected segment pairs represents the behavioral change of an account. We construct two histograms: the \textit{distribution of action behavioral distances} and the \textit{distribution of content behavioral distances}, to represent a user's action and content behavioral changes, respectively. Each histogram is built using 10 bins. We use the value of each bin as a feature, resulting in a total of 20 features that together characterize both aspects of a user's behavioral change (Fig~\ref{fig:overview}). 
In Section~\ref{sec:evaluation}, these 20 features are used to identify inauthentic automation and coordination.

\section{Discriminative Power of Behavioral Change}

Here we examine the distributions of behavioral distances for different classes of accounts across different datasets. Our goal is to show that accounts in the same class tend to exhibit similar distributions of behavioral distances, but different from those of accounts in another class. In the next two subsections we compare automated vs.~human accounts and coordinated vs.~control accounts, respectively. Note that we use the same features (distributions of behavioral distances) for automation and coordination detection. This highlights an important benefit --- the ability of using our proposed method for multiple detection tasks. In this section, we measure the behavioral change of all accounts using the following settings: \textit{sets of 4} segmentation, \textit{cumulative} segment selection, and \textit{cosine} distance. Then, we compare the distributions of action and content behavioral distances for the two classes of accounts. 

\subsection{Automation Behaviors}

We studied automation by investigating the distributions of behavioral change of human and bot accounts labeled by multiple researchers (Table~\ref{tab:automation-datasets}). 
The automated accounts feature a wide variety of suspicious behaviors such as political bots, spam bots, follower bots, and so on.  

\begin{figure*}
    \centering
    
    \begin{subfigure}{0.47\textwidth}
        \includegraphics[width=\linewidth]{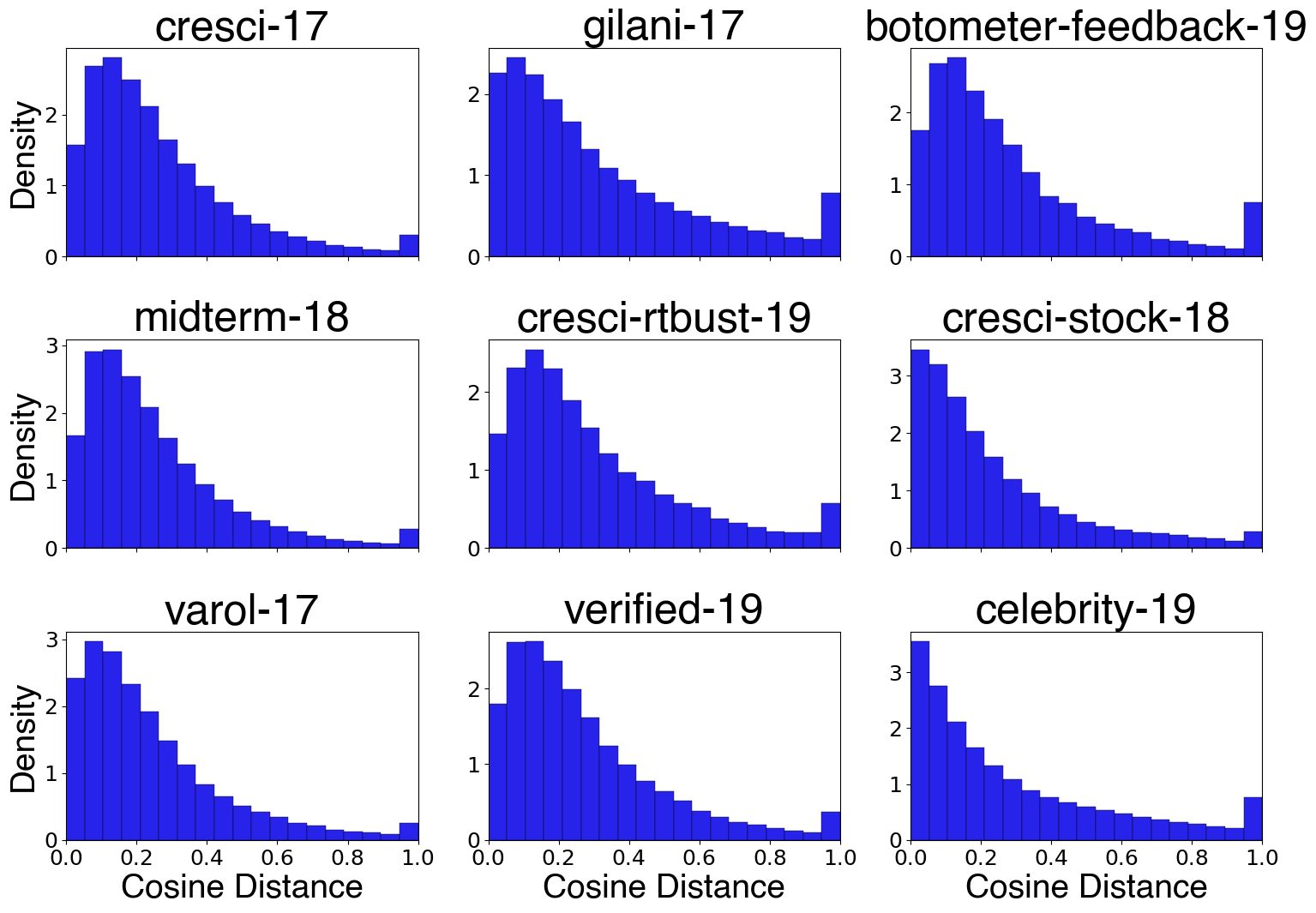}
        \caption{Human -- Action}
        \label{fig:human_action}
    \end{subfigure}
    \hfill
    \begin{subfigure}{0.47\textwidth}
        \includegraphics[width=\linewidth]{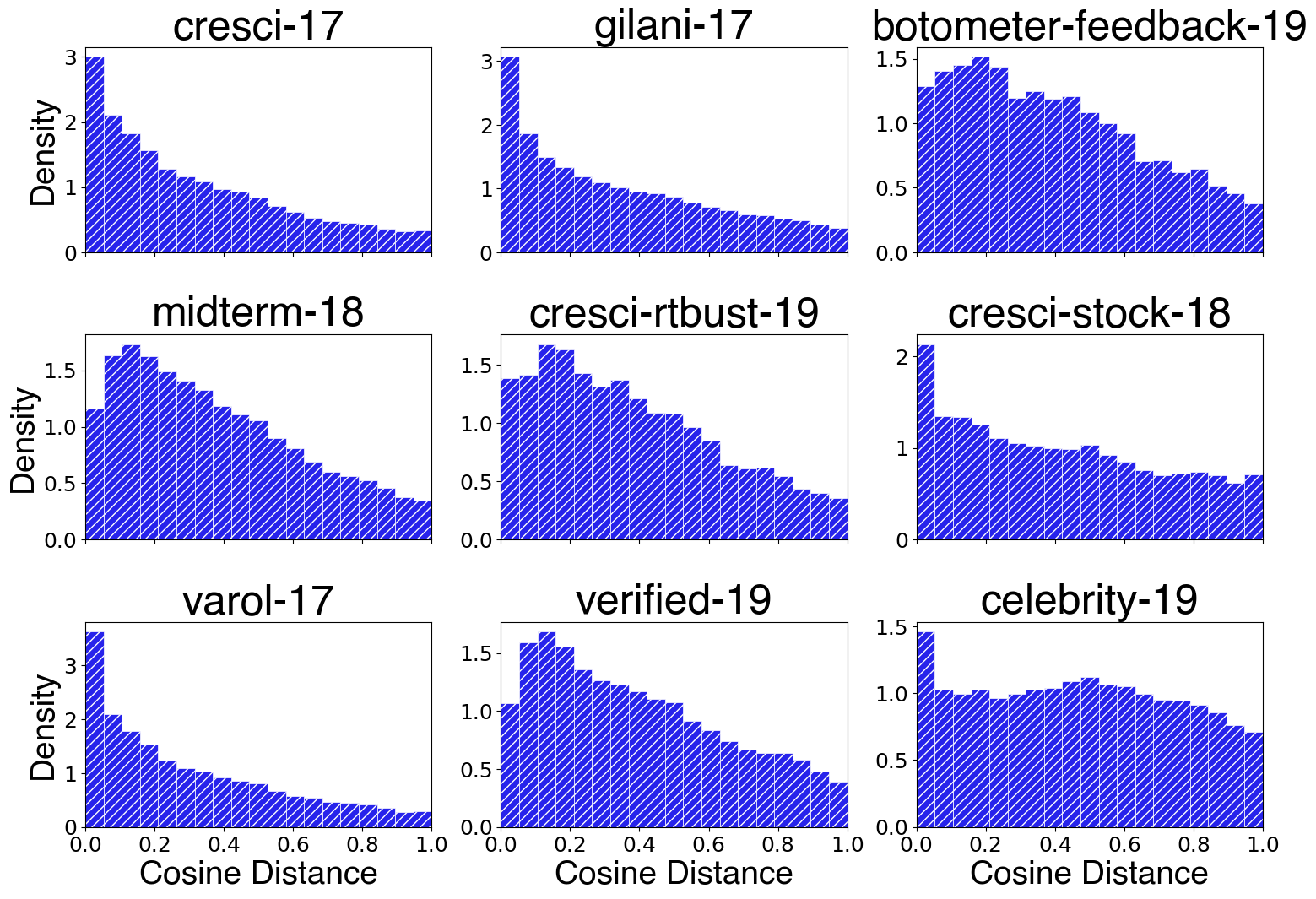}
        \caption{Human -- Content}
        \label{fig:human_content}
    \end{subfigure}
    
    \vspace{0.5em}
    
    \begin{subfigure}{0.47\textwidth}
        \includegraphics[width=\linewidth]{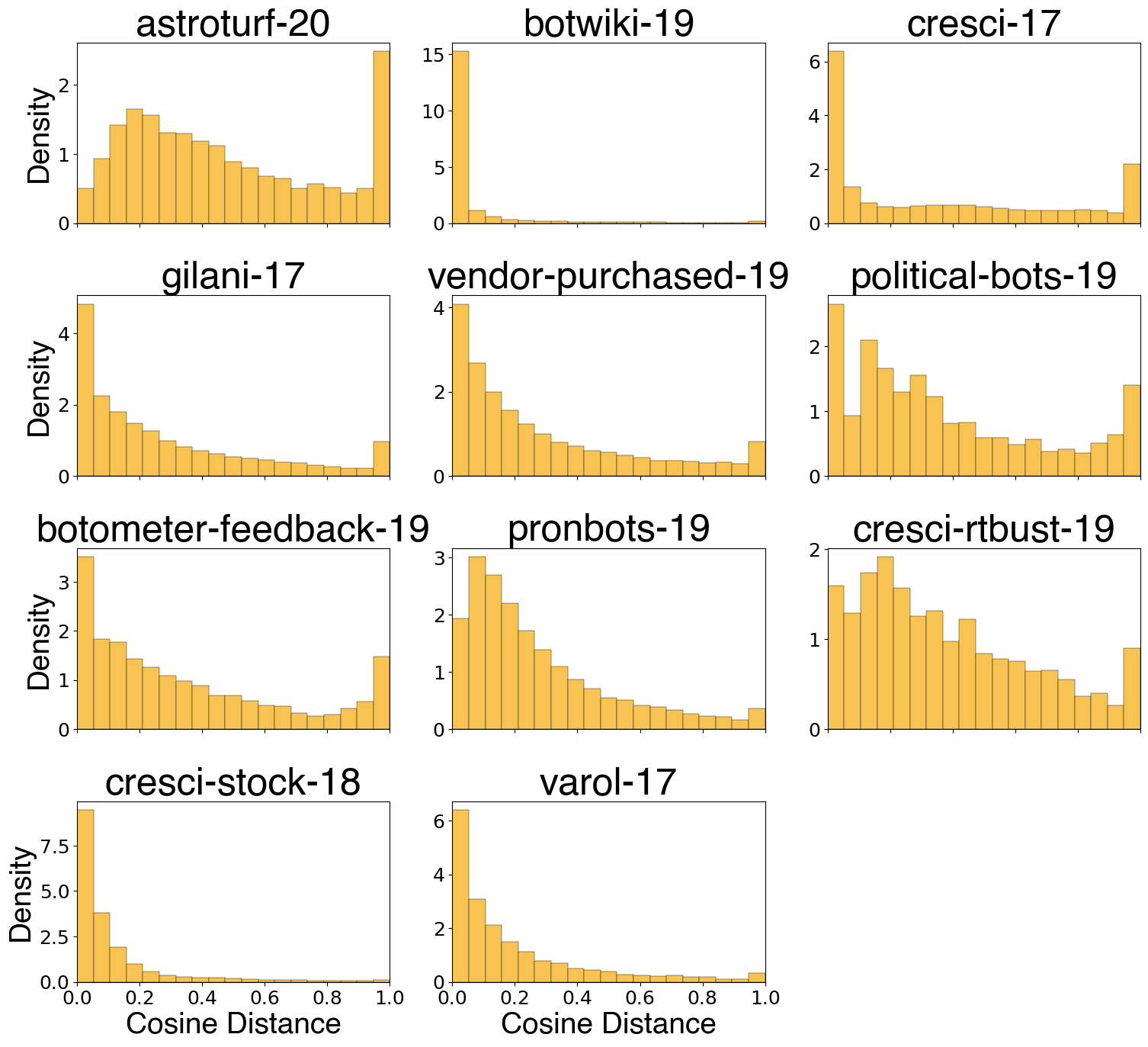}
        \caption{Bot -- Action}
        \label{fig:bot_action}
    \end{subfigure}
    \hfill
    \begin{subfigure}{0.47\textwidth}
        \includegraphics[width=\linewidth]{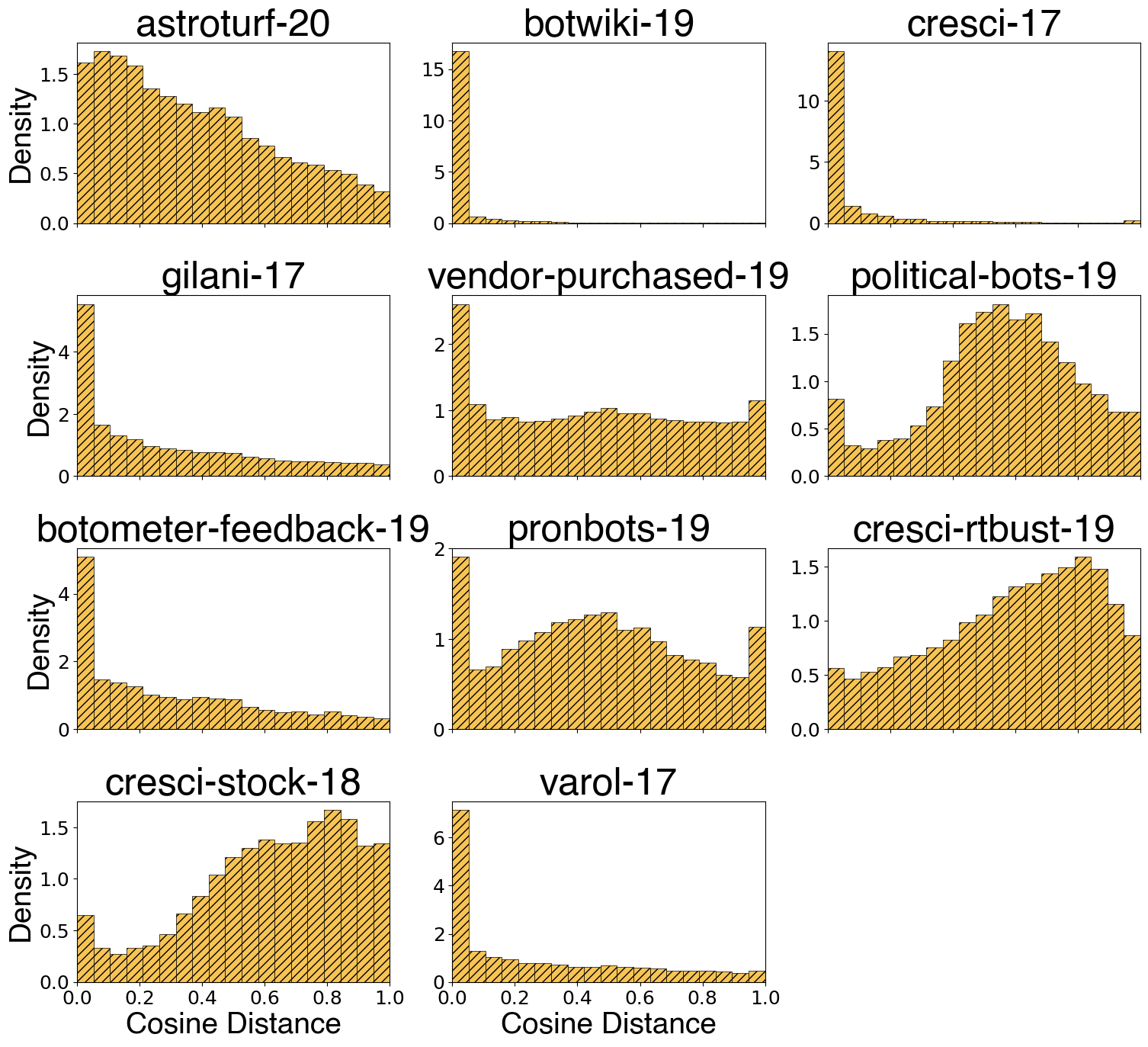}
        \caption{Bot -- Content}
        \label{fig:bot_content}
    \end{subfigure}
     
    \caption{Comparison of the distributions of (a)~action and (b)~content behavioral distances for human accounts, and (c)~action and (d)~content behavioral distances for bot accounts. Colors denote account classes (human in blue and bots in yellow) and patterns denote behavior types (solid for actions and hatched for content).}
    \label{fig:automation_analysis}
\end{figure*}

The distributions of action (Fig.~\ref{fig:human_action}) and content (Fig.~\ref{fig:human_content}) behavioral distances for human accounts across various datasets are similar, with peaks between 0.1 and 0.3, indicating stable behavioral patterns over time. In contrast, the distributions of action (Fig.~\ref{fig:bot_action}) and content (Fig.~\ref{fig:bot_content}) behavioral distances are more varied across datasets of automated accounts.

For example, the peaks for the distributions of content behavioral distances for bot accounts is located at low cosine distance values for some datasets (e.g., \textit{botwiki-19}, \textit{cresci-17}, \textit{gilani-17}) and high cosine distance values for others (e.g., \textit{political-bots-19}, \textit{cresci-rtbust-19}, \textit{cresci-stock-18}).

\begin{table}
\centering
\caption{Datasets used for automation analysis, from the Bot Repository (\url{botometer.osome.iu.edu/bot-repository}).}
\label{tab:automation-datasets}
\small
\begin{tabular}{lcc}
\hline
\textbf{Dataset} & \textbf{\# Bots} & \textbf{\# Humans} \\
\hline
astroturf-20~\cite{sayyadiharikandeh2020detection} & 505 & 0 \\
botometer-feedback-19~\cite{yang2019arming} & 123 & 364 \\
botwiki-19~\cite{yang2020scalable} & 695 & 0 \\
celebrity-19~\cite{yang2019arming} & 0 & 20,911  \\
cresci-17~\cite{cresci2017paradigm} & 5,812 & 2,744 \\
cresci-rtbust-19~\cite{mazza2019rtbust} & 352 & 340 \\
cresci-stock-18~\cite{cresci2019cashtag} & 6,926 & 6,155 \\
gilani-17~\cite{gilani2017depth} & 1,058 & 1,381 \\
midterm-18\cite{yang2020scalable} & 0 & 7,409 \\
political-bots-19~\cite{yang2019arming} & 62 & 0 \\
pronbots-19~\cite{yang2019arming} &  14,867 & 0 \\
varol-17~\cite{varol2017online} & 728 &  1,483 \\
vendor-purchased-19~\cite{yang2019arming} & 928 & 0 \\
verified-19~\cite{yang2020scalable} & 0 & 1,986 \\
\hline
Total &  32,056 & 42,773 \\
\hline
\end{tabular}
\end{table}

\subsection{Coordination Behaviors}
\label{sec:coordination-behaviors}

Similar to automation detection, behavioral change can also help reveal coordinated inauthentic accounts. We studied inauthentic coordinated behaviors using the \textit{AIBot\_fox8}~\cite{anatomyBotnet} and \textit{Information Operations} (IO)~\cite{Seckin_Pote_Nwala_Yin_Luceri_Flammini_Menczer_2025, nwala_flammini_menczer_bloc} datasets. 

The \textit{AIBot\_fox8} dataset~\cite{anatomyBotnet} includes 1,140 coordinated bots and 1,140 genuine human accounts. The bot accounts are from a botnet that spread harmful content such as cryptocurrency scams using ChatGPT for content generation. The bots formed dense reciprocal follow networks and frequently amplified each other's posts. While these accounts may appear normal when examined individually, the key observation is that they exhibit highly similar behavioral patterns, particularly in how they post and interact with content, suggesting inauthentic coordination. To study their behaviors, we compared them to an equal number of genuine human-controlled accounts from the same dataset. These human accounts were sampled evenly from four datasets: \textit{botometer-feedback-19}, \textit{gilani-17}, \textit{midterm-18}, and \textit{varol-17} (Table~\ref{tab:automation-datasets}).

The IO dataset contains \textit{IO accounts} engaged in information operations (\textit{campaigns}) targeting 32 countries, as well as normal (\textit{control}) accounts that discussed similar topics in the same time periods. According to Twitter, an information operation is a form of platform abuse that involves artificial amplification or suppression of information or behavior that manipulates or disrupts the user experience. They often deploy different tactics including automation and coordination. We provide additional information about the IO dataset in Section~\ref{sec:coordination-analysis}.

\begin{figure}
    \centering
    \includegraphics[width=0.47\textwidth]{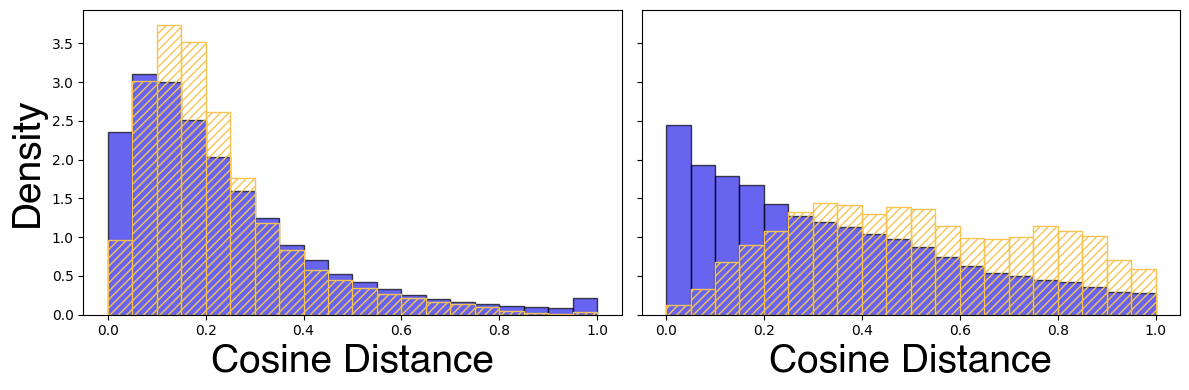}
    \caption{Distributions of action (left) and content (right) behavioral distances for human (blue) and coordinated bot (yellow, hatched) accounts from the \textit{AIBot\_fox8} dataset.}
    \label{fig:fox8-distribution}
\end{figure}

Fig.~\ref{fig:fox8-distribution} shows that human and coordinated bot accounts in the \textit{AIBot\_fox8} dataset have similar distributions of action behavioral distances, with peaks concentrated at low cosine distance values (0.1--0.3). In contrast, their distributions of content behavioral distances reveal stronger differences: while the distances of human accounts clusters more heavily around lower values (0.1--0.3), coordinated bot accounts exhibit a more dispersed distribution extending toward higher distance values, suggesting that these accounts frequently change the content they post. 

\begin{figure}
    \centering
    \begin{subfigure}[b]{0.9\columnwidth}
        \centering
        \includegraphics[width=\textwidth]{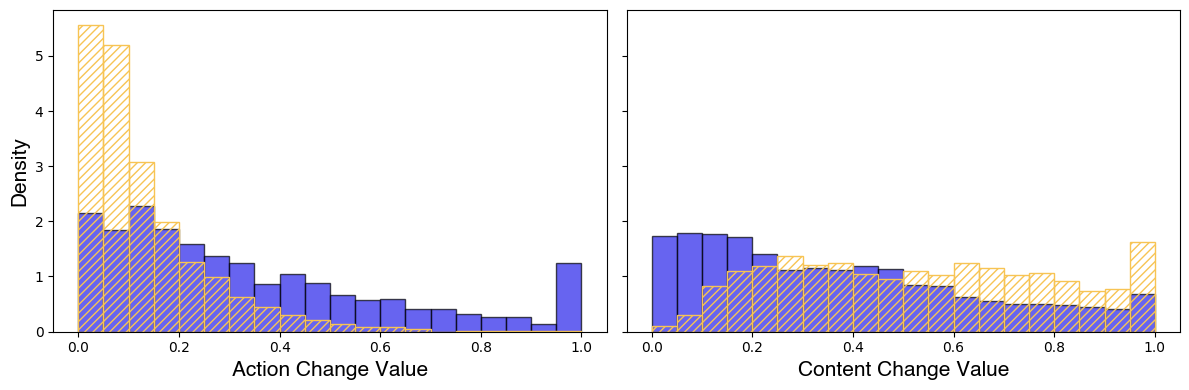}
        \caption{China\_4}
        \label{fig:infoOps-china_4}
    \end{subfigure}

    \begin{subfigure}[b]{0.9\columnwidth}
        \centering
        \includegraphics[width=\textwidth]{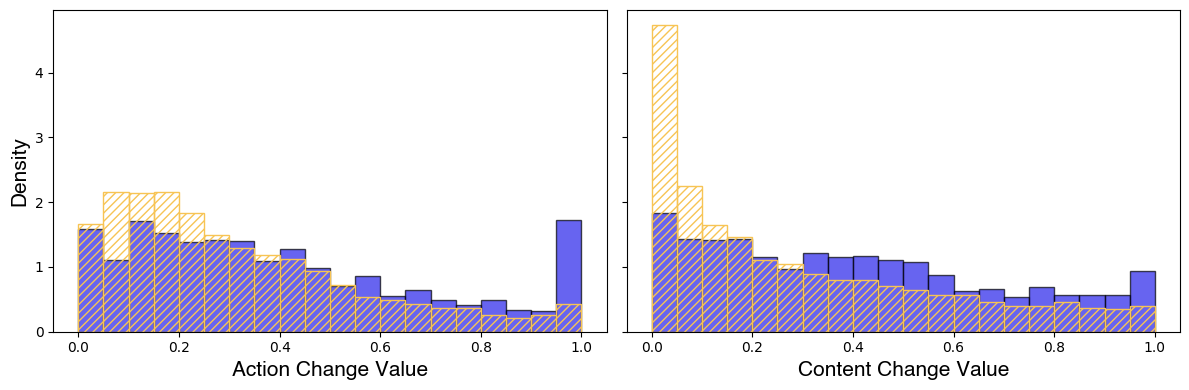}
        \caption{Venezuela\_4}
        \label{fig:infoOps-venezuela_4}
    \end{subfigure}

    \begin{subfigure}[b]{0.9\columnwidth}
        \centering
        \includegraphics[width=\textwidth]{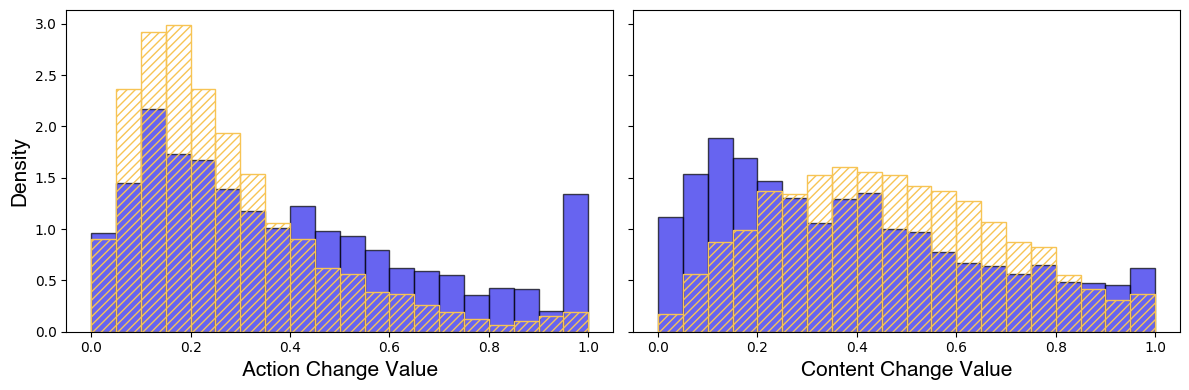}
        \caption{Spain}
        \label{fig:infoOps-spain}
    \end{subfigure}

    \begin{subfigure}[b]{0.9\columnwidth}
        \centering
        \includegraphics[width=\textwidth]{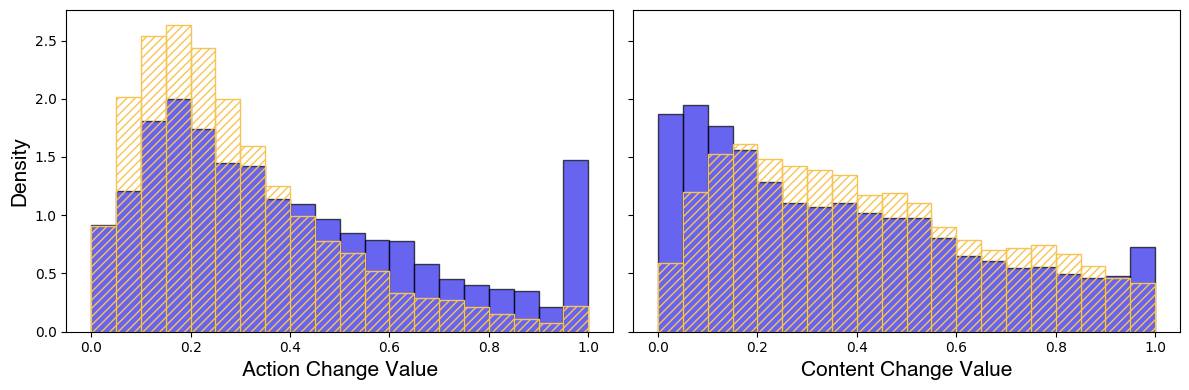}
        \caption{Uganda\_2}
        \label{fig:infoOps-uganda}
    \end{subfigure}

    \caption{Distributions of action (left) and content (right) behavioral distance for control (blue) and IO (yellow, hatched) accounts from four IOs: (a)~\textit{China\_4}, (b)~\textit{Venezuela\_4}, (c)~\textit{Spain}, and (d)~\textit{Uganda\_2}.}
    \label{fig:infoOps-distributions}
\end{figure}

Fig.~\ref{fig:infoOps-distributions} compares the distributions of behavioral distances for \textit{IO} and \textit{control} accounts in the four campaigns with the largest numbers of IO accounts.
In some cases, such as the distributions of action behavioral distances for the \textit{China\_4} campaign and the distributions of content behavioral distances for the \textit{Venezuela\_4} and \textit{Spain} campaigns, we observe marked differences between human and control accounts. 
In other cases, such as action behavioral distances for the \textit{Venezuela\_4} and \textit{Uganda\_2} campaigns, the differences between human and control distributions are smaller, highlighting the difficulty of detecting coordinated accounts --- especially when some campaigns employ actual humans~\cite{ong2018architects}.  

\section{Evaluation}
\label{sec:evaluation}

Let us evaluate the performance of our behavioral change models on two tasks --- identifying automated and coordinated accounts on Twitter. We approach these tasks as supervised machine learning problems and train two separate \textit{Behavior Change} models for each task. As we summarized in Fig.~\ref{fig:overview} and explained in Section~\ref{sec:measuring_change}, for a given task (automation or coordination detection) and a given dataset, we model the behaviors of accounts using BLOC and then measure change through the \textit{distribution of action behavioral distances} and the \textit{distribution of content behavioral distances}. Each distribution is constructed using 10 bins, producing a total of 20 features that jointly characterize the action and content behavioral changes of users. Finally, we used the 20 features to train and evaluate a machine learning model for the task.

\subsection{Automation Detection}

\subsubsection{Dataset} 

We utilized a diverse collection of publicly available labeled Twitter datasets from the bot repository (Table~\ref{tab:automation-datasets}), which has been extensively used by other researchers. These datasets contain accounts labeled as either \textit{bot} (positive class) or \textit{human} (negative class). The following datasets contain the accounts labeled as humans: \textit{verified-19}, \textit{midterm-18}, and \textit{celebrity-19}. The remaining datasets described next cover a diverse range of automation behaviors. The \textit{botwiki-19}, \textit{vendor-purchased-19}, and \textit{botometer-feedback-19} datasets contain bot accounts that self-identify, were purchased, and manually labeled through feedback, respectively. Event-specific datasets includes bot accounts that were active around elections (\textit{political-bots-19} and \textit{astroturf-20}) or stock manipulation (\textit{cresci-stock-18}) events. The \textit{cresci-17} and \textit{cresci-rtbust-19} datasets contain spambots and botnets with similar temporal behaviors.
To maintain data quality, we only included accounts that posted at least 20 tweets. For each selected account, we used up to a maximum of 300 of their most recent tweets. The resulting automation detection evaluation dataset includes a total of 32,056 bot accounts and 42,773 human accounts. 

\subsubsection{Method} 

We systematically explored multiple change settings to build our Behavior Change bot detection model, as described in Section~\ref{sec:measure_change}. We only included users with at least three segments (two distance values). The 20 features from the distributions of action and content behavioral distances were used to train a K-nearest neighbors ($K$-NN) classifier with cosine similarity as the classifier distance metric. We employed five-fold stratified cross-validation. We varied the number of neighbors from $K=1$ to $K=10$, and the performance of each was assessed using the macro-averaged $F_{1}$ score. We then selected the best-performing classifier as the one with the highest macro-$F_{1}$.
We compared our Behavior Change model to Botometer-v4, which was trained and evaluated on the same datasets. Botometer-v4 is one of the most popular state-of-the-art bot detection methods that extracts over 1,000 features from a Twitter account's profile, content, sentiment, social network, and temporal activity~\cite{sayyadiharikandeh2020detection}.

\subsubsection{Results} 

Table~\ref{tab:results} reports the classification performance of our Behavior Change models built using different change settings. 
The combination of the cumulative segment selection method and the compression distance measure consistently produced the strongest results across all segmentation strategies. 
We discuss these results in Section~\ref{sec:discussion}. Table~\ref{tab:automation_benchmark} compares the bot detection classification results of Botometer-v4 with our best-performing Behavior Change model (change setting: \textit{sets-of-4} segmentation, \textit{cumulative} segment selection, and \textit{compression} distance measure). Botometer-v4 slightly outperforms Behavior Change ($F_1$: 0.92 vs. 0.86).

\begin{table}
\centering
\caption{$F_1$ scores of Behavior Change models built using different change settings in the automation detection task. For each segmentation method, the best-performing setting is shown in bold.}
\label{tab:results}
\begin{tabular}{lcc}
\hline
\textbf{Segmentation} & \textbf{Cosine} & \textbf{Compression} \\
\hline
 Sets of four (Adjacent)   & 0.81 & 0.81 \\
 Sets of four (Cumulative) & 0.80 & \textbf{0.86} \\
\hline
Pauses (Adjacent)   & 0.83 & 0.84 \\
Pauses (Cumulative)  & 0.79 & \textbf{0.86} \\
\hline
Weeks (Adjacent)   & 0.77 & 0.76 \\
Weeks (Cumulative) & 0.76 & \textbf{0.80} \\
\hline
\end{tabular}
\end{table}

\begin{table}
\centering
\caption{Precision, recall, and $F_1$ of best-performing Behavior Change bot detection model (change setting: \textit{sets-of-4} segmentation, \textit{cumulative} segment selection, and \textit{compression} distance measure) and \textit{Botometer-v4} using 5-fold cross-validation.}
\begin{tabular}{lcccc}
\hline
\textbf{Model} & \textbf{Precision} & \textbf{Recall} & $\boldsymbol{F}_{1}$ \\
\hline
Behavior Change & 0.86 & 0.86 & 0.86 \\ 
Botometer-v4 & 0.93 & 0.91 & \textbf{0.92} \\ 
\hline
\end{tabular}
\label{tab:automation_benchmark}
\end{table}

\subsection{Coordination Detection}
\label{sec:coordination-analysis}

\subsubsection{Dataset} 

We conducted the coordination analysis using the IO dataset~\cite{Seckin_Pote_Nwala_Yin_Luceri_Flammini_Menczer_2025, nwala_flammini_menczer_bloc} and \textit{AIBot\_fox8} dataset~\cite{anatomyBotnet}, which were introduced in Section~\ref{sec:coordination-behaviors}. We used all accounts in the \textit{AIBot\_fox8} dataset for our evaluation. 
Since the IO dataset includes influence campaigns that spanned months or even years, we focused on the periods when IO accounts were active~\cite{nwala_flammini_menczer_bloc}. 
More specifically, for each campaign in the IO dataset, we added tweets posted by IO accounts in two-week intervals, skipping intervals without IO account activity. 
Once a total of 10 IO accounts were found, the procedure stopped at the end of the year. 
To ensure a fair comparison, we randomly selected the same number of control accounts as IO accounts. This procedure resulted in 32 campaigns suitable for coordination analysis. 

\subsubsection{Method} 

Similarly to automation detection, we selected users with at least two distance values and evaluated multiple Behavior Change models, built using different change settings. Since our automation and coordination detection models use the same underlying Behavior Change system, we used the same 20 features from the distributions of action/content behavioral distances, introduced in Section~\ref{sec:measure_change}. We again employed a $K$-NN classifier with $K=1 \dots 10$ and cosine distance as the classifier distance metric. Here we used leave-one-out cross-validation (LOOCV) to evaluate each model, where each account is held out once as a test instance while the remaining accounts form the training set. Predictions across all LOOCV folds are then aggregated to compute macro-averaged $F_1$ scores, and then we report on the highest $F_1$ across $K$ values. 
We compared our Behavior Change models to three state-of-the-art coordination detection methods~\cite{pacheco2021uncovering} on the same datasets. The \textit{Co-RT} (co-retweet) method identifies coordination by detecting accounts that retweet the same or highly similar sets of posts. The \textit{Hash} (hashtag) method identifies coordinated accounts by using the co-occurrence of hashtags across users, capturing suspicious topical alignment. The \textit{Activity} method relies on temporal synchronization in posting behavior to detect coordination.

\subsubsection{Results} 

Table~\ref{tab:fox8-results} presents the $F_1$ scores of our Behavior Change coordination detection models built using different change settings, for the evaluation conducted on the \textit{AIBot\_fox8} dataset. 
Similarly, Table~\ref{tab:coordination-results-avg} reports mean $F_1$ scores across all campaigns in the IO datasets. 
The control users in the IO dataset have sparse data since a majority of them did not post consistently across the weeks when the IO accounts posted. Consequently, we did not use the segmentation method by weeks in our evaluation. 
Table~\ref{tab:coordination-results_v1} compares $F_1$ scores of our best-performing Behavior Change coordination detection models to the three baselines on both datasets. 
All baselines were evaluated on the same IO datasets to ensure fairness. 
Table~\ref{tab:coordination-results_v2} breaks down the IO results by individual campaigns. We discuss these results in the next section.

\begin{table}
\centering
\caption{$F_1$ scores of Behavior Change models built using different change settings in the coordination detection task, for evaluation conducted on the \textit{AIBot\_fox8} dataset. For each segmentation method, the best-performing setting is shown in bold.}
\label{tab:fox8-results}
\begin{tabular}{lcc}
\toprule
\textbf{Segmentation}        & \textbf{Cosine} & \textbf{Compression} \\
\midrule
Sets of four (Adjacent)      & 0.87 & 0.86 \\
Sets of four (Cumulative)    & \textbf{0.89} & 0.86 \\
\hline
Pauses (Adjacent)            & \textbf{0.94} & 0.93 \\
Pauses (Cumulative)          & 0.90 & 0.88 \\
\hline
Weeks (Adjacent)   & 0.83 & 0.82 \\
Weeks (Cumulative) & \textbf{0.86} & 0.85 \\
\bottomrule
\end{tabular}
\end{table}

\begin{table}
\centering
\caption{Mean $F_1$ scores across all IO campaigns of Behavior Change models built using different change settings in the coordination detection task, for the evaluation conducted on the IO datasets. For each segmentation method, the best-performing setting is shown in bold.}
\label{tab:coordination-results-avg}
\begin{tabular}{lcc}
\toprule
\textbf{Segmentation}        & \textbf{Cosine} & \textbf{Compression} \\
\midrule
Sets of four (Adjacent)      & 0.80 & 0.79 \\
Sets of four (Cumulative)    & 0.73 & \textbf{0.84} \\
\hline
Pauses (Adjacent)            & 0.79 & \textbf{0.88} \\
Pauses (Cumulative)          & 0.73 & 0.86 \\
\bottomrule
\end{tabular}
\end{table}

\begin{table}
\centering
\caption{Precision, recall, and $F_1$ of the best performing Behavior Change coordination detection model and baselines. 
For the \textit{AIBot\_fox8} dataset we use the following change setting: \textit{pauses} segmentation, \textit{adjacent} segment selection, and \textit{cosine} distance measure.
For the IO datasets we use \textit{pauses} segmentation, \textit{adjacent} segment selection, and \textit{compression} distance measure, and report mean $F_1$ across IO campaigns.}
\begin{tabular}{llccc}
\hline
\textbf{Dataset} & \textbf{Model} & \textbf{Precision} & \textbf{Recall} & $\boldsymbol{F}_{1}$ \\
\hline
\multirow{4}{*}{AIBot\_fox8} & Behavior Change & 0.94 &  0.94 &  0.94 \\
& Co-RT  & 0.99 & \textbf{1.00} &  \textbf{0.99} \\
& Hashtag  & \textbf{1.00} & 0.21 & 0.34  \\
& Activity  & 0.99 & \textbf{1.00} &  \textbf{0.99} \\
\hline
\multirow{4}{*}{IO} & Behavior Change & \textbf{0.89} & 0.88  &  \textbf{0.88}  \\
& Co-RT  & 0.43 & 0.71 & 0.53  \\
& Hashtag  & 0.46  & 0.81 & 0.56  \\
& Activity  & 0.64 & \textbf{0.95} & 0.76  \\
\hline
\end{tabular}
\label{tab:coordination-results_v1}
\end{table}

\begin{table}
\caption{$F_1$ scores of the best-performing Behavior Change coordination detection model compared to three baselines, for each IO campaign. The Behavior Change model was built using the following change setting: \textit{pauses} segmentation, \textit{adjacent} segment selection method, and \textit{compression} distance measure. The best result for each row is shown in bold.}
\centering
\begin{tabular}{l p{1cm} p{1.5cm} p{1cm} p{1cm}}
\hline
\textbf{Dataset} & \textbf{Behav. Change}  & \textbf{Hashtag} & \textbf{Co-RT} & \textbf{Activity} \\
\hline
Qatar & 0.71 & 0.67 & 0.73 & \textbf{0.81} \\
Iran\_5 & 0.74 & 0.64 & 0.69 & \textbf{0.79} \\
Egypt\_UAE & 0.74 & 0.00 & 0.00 & \textbf{0.76} \\
Russia\_1 & 0.74 & 0.62 & \textbf{0.71} & 0.50 \\
Russia\_3 & 0.78 & 0.67 & 0.69 & \textbf{0.80} \\
Iran\_3 & \textbf{0.78} & 0.33 & 0.00 & 0.64 \\
Spain & 0.81 & 0.92 & \textbf{0.98} & 0.96 \\
Ecuador & \textbf{0.82} & 0.00 & 0.64 & 0.71 \\
Armenia & \textbf{0.83} & 0.67 & 0.69 & 0.78 \\
China\_5 & \textbf{0.85} & 0.66 & 0.67 & 0.83 \\
Iran\_1 & \textbf{0.86} & 0.63 & 0.00 & 0.76 \\
Uganda\_2 & 0.87 & 0.76 & \textbf{0.99} & 0.75 \\
Venezuela\_4 & 0.87 & 0.95 & \textbf{0.95} & 0.84 \\
Venezuela\_1 & \textbf{0.87} & 0.61 & 0.78 & 0.66 \\
Iran\_4 & \textbf{0.89} & 0.67 & 0.71 & 0.77 \\
Russia\_4 & \textbf{0.90} & 0.00 & 0.00 & 0.77 \\
Iran\_2 & \textbf{0.91} & 0.58 & 0.67 & 0.74 \\
Uganda\_1 & \textbf{0.91} & 0.67 & 0.69 & 0.67 \\
Catalonia & \textbf{0.92} & 0.67 & 0.67 & 0.79 \\
UAE & \textbf{0.92} & 0.00 & 0.00 & 0.72 \\
Mexico\_1 & \textbf{0.92} & 0.66 & 0.63 & 0.67 \\
China\_2 & \textbf{0.92} & 0.63 & 0.00 & 0.69 \\
Ghana\_Nigeria & \textbf{0.92} & 0.67 & 0.66 & 0.75 \\
Bangladesh & \textbf{0.93} & 0.56 & 0.00 & 0.80 \\
Venezuela\_5 & \textbf{0.93} & 0.66 & 0.65 & 0.65 \\
China\_4 & 0.93 & 0.97 & \textbf{0.99} & 0.98 \\
Iran\_7 & \textbf{0.93} & 0.64 & 0.61 & 0.91 \\
Thailand & \textbf{0.94} & 0.67 & 0.79 & 0.72 \\
Venezuela\_3 & \textbf{0.96} & 0.00 & 0.56 & 0.72 \\
Iran\_6 & \textbf{0.97} & 0.67 & 0.69 & 0.75 \\
China\_1 & \textbf{1.00} & 0.59 & 0.00 & 0.65 \\
Venezuela\_2 & \textbf{1.00} & 0.58 & 0.00 & 0.84 \\
\hline
\textbf{Average} & \textbf{0.88} & 0.56 & 0.53 & 0.76 \\
\hline
\end{tabular}
\label{tab:coordination-results_v2}
\end{table}

\section{Discussion}
\label{sec:discussion}

We address two major problems of existing social media manipulation detection systems. First, most current methods are designed for specific tasks, such as detecting spam bots, identifying coordinated accounts, or finding hacked accounts. Consequently, they do not generalize well to other tasks. Second, suspicious users often change their behaviors over time to evade detection, which makes these systems less reliable in the long run. To overcome these challenges, we introduced a new approach that uses behavioral change itself as a signal for detecting suspicious activity. This idea can be applied to detect multiple suspicious behaviors. In our work, we focused on two examples: detecting automation and detecting inauthentic coordination. 

In addition to its generality, our proposed method is also flexible. Specifically, one can replace BLOC, which we used to represent user behaviors, with another representation that captures actions and/or content behaviors and then apply a suitable technique to measure distances between behavioral representations. Once distance values are obtained, they can be used to create features to train a detection model. Even though our investigations used Twitter data, our framework can be applied to other social media platforms, provided that user behavior can be represented in a similar way.

We found that genuine human accounts exhibit neither very repetitive behaviors nor very drastic changes, while the behavioral changes of bot accounts deviate from those of genuine humans (Fig.~\ref{fig:automation_analysis}). Some exhibit unusually small change, suggesting automated behavior, while others display large changes, possibly when they are repurposed or to avoid detection.
We evaluated Behavior Change models for bot detection with different change settings (Table~\ref{tab:results}). The compression distance measure may perform well because it preserves the sequential order of user actions when comparing segments, whereas cosine distance discards temporal information. The cumulative segmentation method, capturing both short-term and long-term changes in behavior, also performed effectively. Segmenting by either sets-of-4 or pauses consistently outperformed segmenting by weeks. The latter could result in comparing pairs of segments with significantly different lengths, which could happen if a user actively posted content in one week but not in the following week. 
Although Botometer-V4 achieved higher overall $F_1$ (Table~\ref{tab:automation_benchmark}) than our Behavior Change model, it relies on multiple classes of features extracted from a user's profile, content, sentiment, network, and temporal attributes to accomplish this result. In contrast, our Behavior Change model only uses features extracted from a user's action and content behaviors. These results show that behavioral change alone provides valuable insight for identifying automated accounts.

We observed that coordinated accounts tend to display patterns of behavioral change that are similar to each other and different from those of control accounts (Fig.~\ref{fig:fox8-distribution} and \ref{fig:infoOps-distributions}). 
Our Behavior Change models performed consistently well at detecting coordination across different change settings (Tables~\ref{tab:fox8-results} and~\ref{tab:coordination-results-avg}) and in both evaluation datasets (Table~\ref{tab:coordination-results_v1}). 

In particular, Behavior Change outperformed the baselines on average across IO campaigns and in most individual campaigns (Table~\ref{tab:coordination-results_v2}). 
Pause-based segmentation, which captures how user behavior evolves across discrete activity sessions, produced the strongest results for both evaluation datasets. 
The adjacent segmentation method worked better here, while different similarity measures yielded higher $F_1$ in different datasets. 
Among the baselines, Co-RT and Activity performed best on the \textit{AIBot\_fox8} dataset. These coordinated bots share similar content, making them easier for the Co-RT method to detect. Additionally, these bots post on timelines that differ noticeably from human accounts in the dataset, enabling the Activity method to differentiate them more clearly. However, the performance of both baselines did not translate to the IO datasets, which involves accounts employing a wider range of tactics.

We focused exclusively on user actions and content, but we recognize that there are other important aspects of user behavior that could be explored in future work. This includes capturing changes in topics and patterns of interactions (e.g., following accounts or reacting to posts). Considering these additional aspects of behaviors could provide a fuller picture of behavioral change and help detect an even wider range of suspicious behaviors. 

\section{Acknowledgments}
The authors acknowledge William \& Mary Research Computing for providing computational resources and technical support that have contributed to the results reported within this paper. 

\printbibliography

@article{fake_news_spreader,
    author = {Sahar Baribi-Bartov  and Briony Swire-Thompson  and Nir Grinberg },
    title = {Supersharers of fake news on Twitter},
    journal = {Science},
    volume = {384},
    number = {6699},
    pages = {979-982},
    year = {2024},
    doi = {10.1126/science.adl4435},
    URL = {https://www.science.org/doi/abs/10.1126/science.adl4435},
}

@article{2016_presidential_election,
author = {Nir Grinberg  and Kenneth Joseph  and Lisa Friedland  and Briony Swire-Thompson  and David Lazer },
title = {Fake news on Twitter during the 2016 U.S. presidential election},
journal = {Science},
volume = {363},
number = {6425},
pages = {374-378},
year = {2019},
doi = {10.1126/science.aau2706},
}

@article{bessi2016social,
  title={Social bots distort the 2016 US Presidential election online discussion},
  author={Bessi, Alessandro and Ferrara, Emilio},
  journal={First monday},
  volume={21},
  number={11-7},
  year={2016}
}

@article{deverna2025modeling,
  title={Modeling the amplification of epidemic spread by individuals exposed to misinformation on social media},
  author={DeVerna, Matthew R and Pierri, Francesco and Ahn, Yong-Yeol and Fortunato, Santo and Flammini, Alessandro and Menczer, Filippo},
  journal={npj Complexity},
  volume={2},
  number={11},
  year={2025},
  url={https://doi.org/10.1038/s44260-025-00038-y}
}

@article{decade_of_social_bots,
author = {Cresci, Stefano},
title = {A decade of social bot detection},
year = {2020},
issue_date = {October 2020},
publisher = {Association for Computing Machinery},
address = {New York, NY, USA},
volume = {63},
number = {10},
issn = {0001-0782},
url = {https://doi.org/10.1145/3409116},
doi = {10.1145/3409116},
journal = {Commun. ACM},
month = sep,
pages = {72–83},
numpages = {12}
}

@article{cresci2023demystifying,
    author = {Stefano Cresci and Kai-Cheng Yang and Angelo Spognardi and Roberto Di Pietro and Filippo Menczer and Marinella Petrocchi},
    title ={Demystifying Misconceptions in Social Bots Research},
    journal = {Social Science Computer Review},
    volume = {0},
    number = {0},
    pages = {08944393251376707},
    year = {2025},
    doi = {10.1177/08944393251376707},
    URL = {https://doi.org/10.1177/08944393251376707},
}

@misc{yang2023social,
      title={Social Bots: Detection and Challenges}, 
      author={Kai-Cheng Yang and Onur Varol and Alexander C. Nwala and Mohsen Sayyadiharikandeh and Emilio Ferrara and Alessandro Flammini and Filippo Menczer},
      year={2023},
      archivePrefix={arXiv},
      primaryClass={cs.SI},
      url={https://arxiv.org/abs/2312.17423}, 
}

@INPROCEEDINGS{dickerson2014using,
  author={Dickerson, John P. and Kagan, Vadim and Subrahmanian, V.S.},
  booktitle={2014 IEEE/ACM International Conference on Advances in Social Networks Analysis and Mining (ASONAM)}, 
  title={Using sentiment to detect bots on Twitter: Are humans more opinionated than bots?}, 
  year={2014},
  volume={},
  number={},
  pages={620--627},
  doi={10.1109/ASONAM.2014.6921650}
}

@article{pacheco2021uncovering, 
    title={Uncovering Coordinated Networks on Social Media: Methods and Case Studies}, 
    volume={15}, 
    url={https://ojs.aaai.org/index.php/ICWSM/article/view/18075}, 
    DOI={10.1609/icwsm.v15i1.18075}, 
    number={1},
    journal={Proceedings of the International AAAI Conference on Web and Social Media}, 
    author={Pacheco, Diogo and Hui, Pik-Mai and Torres-Lugo, Christopher and Truong, Bao Tran and Flammini, Alessandro and Menczer, Filippo}, 
    year={2021}, 
    month={May}, 
    pages={455-466} 
}

@article{ali2023real,
  title={Real-Time Spammers Detection Based on Metadata Features with Machine Learning.},
  author={Ali, Adnan and Li, Jinlong and Chen, Huanhuan and Bhatti, Uzair Aslam and Khan, Asad},
  journal={Intelligent Automation \& Soft Computing},
  volume={38},
  number={3},
  year={2023}
}

@article{rodriguez2020one,
    title = {A one-class classification approach for bot detection on Twitter},
    journal = {Computers \& Security},
    volume = {91},
    pages = {101715},
    year = {2020},
    issn = {0167-4048},
    doi = {https://doi.org/10.1016/j.cose.2020.101715},
    url = {https://www.sciencedirect.com/science/article/pii/S0167404820300031},
    author = {Jorge Rodríguez-Ruiz and Javier Israel Mata-Sánchez and Raúl Monroy and Octavio Loyola-González and Armando López-Cuevas},
}

@article{nwala_flammini_menczer_bloc,
  title={A language framework for modeling social media account behavior},
  author={Nwala, Alexander C and Flammini, Alessandro and Menczer, Filippo},
  journal={EPJ Data Science},
  volume={12},
  number={1},
  pages={33},
  year={2023},
  publisher={Springer Berlin Heidelberg}
}

@inproceedings{cresci2017paradigm,
author = {Cresci, Stefano and Di Pietro, Roberto and Petrocchi, Marinella and Spognardi, Angelo and Tesconi, Maurizio},
title = {The Paradigm-Shift of Social Spambots: Evidence, Theories, and Tools for the Arms Race},
year = {2017},
url = {https://doi.org/10.1145/3041021.3055135},
doi = {10.1145/3041021.3055135},
booktitle = {Proceedings of the 26th International Conference on WWW Companion},
pages = {963–-972},
numpages = {10},
}

@article{ferrara2016rise,
author = {Ferrara, Emilio and Varol, Onur and Davis, Clayton and Menczer, Filippo and Flammini, Alessandro},
title = {The rise of social bots},
year = {2016},
issue_date = {July 2016},
publisher = {Association for Computing Machinery},
address = {New York, NY, USA},
volume = {59},
number = {7},
issn = {0001-0782},
url = {https://doi.org/10.1145/2818717},
doi = {10.1145/2818717},
journal = {Commun. ACM},
month = jun,
pages = {96–104},
numpages = {9}
}

@inproceedings{cao2014uncovering,
author = {Cao, Qiang and Yang, Xiaowei and Yu, Jieqi and Palow, Christopher},
title = {Uncovering Large Groups of Active Malicious Accounts in Online Social Networks},
year = {2014},
isbn = {9781450329576},
url = {https://doi.org/10.1145/2660267.2660269},
doi = {10.1145/2660267.2660269},
booktitle = {Proceedings of the 2014 ACM SIGSAC Conference on Computer and Communications Security},
pages = {477–488},
numpages = {12},
series = {CCS '14}
}

@article{costa2017modeling,
author = {Costa, Alceu Ferraz and Yamaguchi, Yuto and Traina, Agma Juci Machado and Jr., Caetano Traina and Faloutsos, Christos},
title = {Modeling Temporal Activity to Detect Anomalous Behavior in Social Media},
year = {2017},
issue_date = {November 2017},
publisher = {Association for Computing Machinery},
address = {New York, NY, USA},
volume = {11},
number = {4},
url = {https://doi.org/10.1145/3064884},
doi = {10.1145/3064884},
journal = {ACM Trans. Knowl. Discov. Data},
articleno = {49}
}

@INPROCEEDINGS{mannocci2022mulbot,
  author={Mannocci, Lorenzo and Cresci, Stefano and Monreale, Anna and Vakali, Athina and Tesconi, Maurizio},
  booktitle={2022 IEEE International Conference on Big Data (Big Data)}, 
  title={MulBot: Unsupervised Bot Detection Based on Multivariate Time Series}, 
  year={2022},
  volume={},
  number={},
  pages={1485-1494},
  doi={10.1109/BigData55660.2022.10020363}
}

@inproceedings{wu2023botshape, 
    series={ICAITA 2023},
   title={Botshape: A Novel Social Bots Detection Approach via Behavioral Patterns},
   url={http://dx.doi.org/10.5121/csit.2023.130604},
   DOI={10.5121/csit.2023.130604},
   booktitle={Advanced Information Technologies and Applications},
   publisher={Academy and Industry Research Collaboration Center (AIRCC)},
   author={Wu, Jun and Ye, Xuesong and Mou, Chengjie},
   year={2023},
   month=mar, pages={45–60},
   collection={ICAITA 2023} 
}

@article{pedersen2023detecting,
  title={Detecting bots with temporal logic},
  author={Pedersen, Mina Young and Slavkovik, Marija and Smets, Sonja},
  journal={Synthese},
  volume={202},
  number={3},
  pages={79},
  year={2023},
  publisher={Springer}
}

@ARTICLE{cresci2016dna,
  author={Cresci, Stefano and Di Pietro, Roberto and Petrocchi, Marinella and Spognardi, Angelo and Tesconi, Maurizio},
  journal={IEEE Intelligent Systems}, 
  title={DNA-Inspired Online Behavioral Modeling and Its Application to Spambot Detection}, 
  year={2016},
  volume={31},
  number={5},
  pages={58-64},
  doi={10.1109/MIS.2016.29}
}

@article{cresci2019cashtag,
author = {Cresci, Stefano and Lillo, Fabrizio and Regoli, Daniele and Tardelli, Serena and Tesconi, Maurizio},
title = {Cashtag Piggybacking: Uncovering Spam and Bot Activity in Stock Microblogs on Twitter},
year = {2019},
issue_date = {May 2019},
publisher = {Association for Computing Machinery},
address = {New York, NY, USA},
volume = {13},
number = {2},
issn = {1559-1131},
url = {https://doi.org/10.1145/3313184},
doi = {10.1145/3313184},
journal = {ACM Trans. Web},
month = apr,
articleno = {11},
numpages = {27}
}

@misc{farazmanesh2022compromised,
      title={Compromised account detection using authorship verification: a novel approach}, 
      author={Forough Farazmanesh and Fateme Foroutan and Amir Jalaly Bidgoly},
      year={2022},
      archivePrefix={arXiv},
      primaryClass={cs.CR},
      url={https://arxiv.org/abs/2206.03581}, 
}

@inproceedings{nauta2017detecting,
  title={Detecting hacked twitter accounts based on behavioural change},
  author={Nauta, Meike and Habib, Mena Badieh and van Keulen, Maurice},
  booktitle={13th International Conference on Web Information Systems and Technologies (WEBIST)},
  pages={19--31},
  year={2017},
}

@article{yen2021detecting,
  title={Detecting compromised social network accounts using deep learning for behavior and text analyses},
  author={Yen, Steven and Moh, Melody and Moh, Teng-Sheng},
  journal={International Journal of Cloud Applications and Computing},
  volume={11},
  number={2},
  pages={1--13},
  year={2021},
}

@article{diaz2025survey,
  title={A survey of textual cyber abuse detection using cutting-edge language models and large language models},
  author={Diaz-Garcia, Jose A and Carvalho, Joao Paulo},
  journal={arXiv preprint arXiv:2501.05443},
  year={2025}
}

@article{varol2017online, 
title={Online Human-Bot Interactions: Detection, Estimation, and Characterization}, 
volume={11}, 
url={https://ojs.aaai.org/index.php/ICWSM/article/view/14871},
DOI={10.1609/icwsm.v11i1.14871},
number={1}, 
journal={Proceedings of the International AAAI Conference on Web and Social Media}, 
author={Varol, Onur and Ferrara, Emilio and Davis, Clayton and Menczer, Filippo and Flammini, Alessandro}, 
year={2017}, 
pages={280-289} 
}

@article{anatomyBotnet, 
title={Anatomy of an AI-powered malicious social botnet},
volume={4}, url={https://journalqd.org/article/view/5848}, 
DOI={10.51685/jqd.2024.icwsm.7},
journal={Journal of Quantitative Description: Digital Media}, 
author={Yang, Kai-Cheng and Menczer, Filippo}, 
year={2024} 
}

@article{socioLinguisticCia, 
title={Socio-Linguistic Characteristics of Coordinated Inauthentic Accounts}, 
volume={18}, 
url={https://ojs.aaai.org/index.php/ICWSM/article/view/31305}, 
DOI={10.1609/icwsm.v18i1.31305}, 
number={1}, 
journal={Proceedings of the International AAAI Conference on Web and Social Media}, 
author={Burghardt, Keith and Rao, Ashwin and Chochlakis, Georgios and Sabyasachee, Baruah and Guo, Siyi and He, Zihao and Rojecki, Andrew and Narayanan, Shrikanth and Lerman, Kristina}, 
year={2024}, month={May}, 
pages={164-176} 
}

@inproceedings {thomas2013trafficking,
	author = {Kurt Thomas and Damon McCoy and Chris Grier and Alek Kolcz and Vern Paxson},
	title = {{Trafficking} Fraudulent Accounts: The Role of the Underground Market in Twitter Spam and Abuse},
	booktitle = {22nd USENIX Security Symposium},
	year = {2013},
	pages = {195--210},
	url = {https://www.usenix.org/conference/usenixsecurity13/technical-sessions/paper/thomas},
}

@misc{inuwa2018lexical,
      title={Lexical analysis of automated accounts on Twitter}, 
      author={Isa Inuwa-Dutse and Bello Shehu Bello and Ioannis Korkontzelos},
      year={2018},
      archivePrefix={arXiv},
      primaryClass={cs.SI},
      url={https://arxiv.org/abs/1812.07947}, 
}

@misc{gilani2017depth,
      title={An in-depth characterisation of Bots and Humans on Twitter}, 
      author={Zafar Gilani and Reza Farahbakhsh and Gareth Tyson and Liang Wang and Jon Crowcroft},
      year={2017},
      archivePrefix={arXiv},
      primaryClass={cs.SI},
      url={https://arxiv.org/abs/1704.01508}, 
}

@ARTICLE{cilibrasi2005clustering,
  author={Cilibrasi, R. and Vitanyi, P.M.B.},
  journal={IEEE Transactions on Information Theory}, 
  title={Clustering by compression}, 
  year={2005},
  volume={51},
  number={4},
  pages={1523-1545},
}

@article{yang2020scalable, 
title={Scalable and Generalizable Social Bot Detection through Data Selection}, 
volume={34}, 
url={https://ojs.aaai.org/index.php/AAAI/article/view/5460}, 
DOI={10.1609/aaai.v34i01.5460},
number={01}, 
journal={Proceedings of the AAAI Conference on Artificial Intelligence}, 
author={Yang, Kai-Cheng and Varol, Onur and Hui, Pik-Mai and Menczer, Filippo}, 
year={2020}, 
month={Apr.}, 
pages={1096-1103} 
}

@article{yang2019arming,
author = {Yang, Kai-Cheng and Varol, Onur and Davis, Clayton A. and Ferrara, Emilio and Flammini, Alessandro and Menczer, Filippo},
title = {Arming the public with artificial intelligence to counter social bots},
journal = {Human Behavior and Emerging Technologies},
volume = {1},
number = {1},
pages = {48-61},
doi = {10.1002/hbe2.115},
year = {2019}
}

@inproceedings{mazza2019rtbust,
author = {Mazza, Michele and Cresci, Stefano and Avvenuti, Marco and Quattrociocchi, Walter and Tesconi, Maurizio},
title = {RTbust: Exploiting Temporal Patterns for Botnet Detection on Twitter},
year = {2019},
url = {https://doi.org/10.1145/3292522.3326015},
doi = {10.1145/3292522.3326015},
booktitle = {Proceedings of the 10th ACM Conference on Web Science},
pages = {183–192},
numpages = {10},
series = {WebSci '19}
}

@article{chu2012detecting,
  title={Detecting automation of twitter accounts: Are you a human, bot, or cyborg?},
  author={Chu, Zi and Gianvecchio, Steven and Wang, Haining and Jajodia, Sushil},
  journal={IEEE Transactions on dependable and secure computing},
  volume={9},
  number={6},
  pages={811--824},
  year={2012},
  publisher={IEEE}
}

@article{Seckin_Pote_Nwala_Yin_Luceri_Flammini_Menczer_2025, 
    title={Labeled Datasets for Research on Information Operations}, volume={19}, url={https://ojs.aaai.org/index.php/ICWSM/article/view/35958}, DOI={10.1609/icwsm.v19i1.35958}, 
    number={1}, 
    journal={Proceedings of the International AAAI Conference on Web and Social Media}, 
    author={Seckin, Ozgur Can and Pote, Manita and Nwala, Alexander C and Yin, Lake and Luceri, Luca and Flammini, Alessandro and Menczer, Filippo}, 
    year={2025},
    month={Jun.}, 
    pages={2567-2574} 
}

@inproceedings{sayyadiharikandeh2020detection,
author = {Sayyadiharikandeh, Mohsen and Varol, Onur and Yang, Kai-Cheng and Flammini, Alessandro and Menczer, Filippo},
title = {Detection of Novel Social Bots by Ensembles of Specialized Classifiers},
year = {2020},
url = {https://doi.org/10.1145/3340531.3412698},
doi = {10.1145/3340531.3412698},
booktitle = {Proceedings of the 29th ACM International Conference on Information \& Knowledge Management},
pages = {2725–2732},
numpages = {8},
series = {CIKM}
}

@article{wack2025generative,
  title={Generative propaganda: Evidence of AI’s impact from a state-backed disinformation campaign},
  author={Wack, Morgan and Ehrett, Carl and Linvill, Darren and Warren, Patrick},
  journal={PNAS nexus},
  volume={4},
  number={4},
  pages={pgaf083},
  year={2025},
  publisher={Oxford University Press US}
}

@article{smith2025unsupervised,
  title={Unsupervised detection of coordinated information operations in the wild},
  author={Smith, D Hudson and Ehrett, Carl and Warren, Patrick},
  journal={EPJ Data Science},
  volume={14},
  number={1},
  pages={26},
  year={2025},
  publisher={Springer}
}

@misc{weber2022temporalc,
      title={Temporal Nuances of Coordination Network Semantics}, 
      author={Derek Weber and Lucia Falzon},
      year={2022},
      archivePrefix={arXiv},
      primaryClass={cs.SI},
      url={https://arxiv.org/abs/2107.02588}, 
}

@misc{ong2018architects,
  title={Architects of networked disinformation: Behind the scenes of troll accounts and fake news production in the Philippines},
  author={Ong, Jonathan Corpus and Caba{\~n}es, Jason Vincent},
  year={2018},
  howpublished={Newton Tech4Dev Network},
  url={https://doi.org//10.7275/2cq4-5396}
}

@article{Cresci_Petrocchi_Spognardi_Tognazzi_2021, 
    title={The coming age of adversarial social bot detection}, 
    volume={26}, 
    url={https://firstmonday.org/ojs/index.php/fm/article/view/11474}, 
    DOI={10.5210/fm.v26i7.11474}, 
    number={7}, 
    journal={First Monday}, 
    author={Cresci, Stefano and Petrocchi, Marinella and Spognardi, Angelo and Tognazzi, Stefano}, 
    year={2021}, 
    month={May} 
}

@misc{elmas2022misleadingrepurposingtwitter,
      title={Misleading Repurposing on Twitter}, 
      author={Tuğrulcan Elmas and Rebekah Overdorf and Karl Aberer},
      year={2022},
      archivePrefix={arXiv},
      primaryClass={cs.SI},
      url={https://arxiv.org/abs/2010.10600}, 
}

\end{document}